\documentclass[a4paper,11pt]{article}

\usepackage{jheppub} 
\usepackage{jheppub}
\usepackage{subfigure}
\usepackage{subfigure}
\usepackage{ascmac}
\usepackage{amsthm}
\usepackage{amsmath}
\usepackage{tikz}
\usepackage{pifont}
\usepackage{booktabs}
\usetikzlibrary{arrows.meta}
\usetikzlibrary{patterns}
\usetikzlibrary{decorations.pathmorphing}
\usetikzlibrary{decorations.text}
\tikzset{>={Latex[width=2mm,length=2mm]}}

\preprint{}

\title{\boldmath
Reconstructing black hole exteriors and interiors using entanglement and complexity
}

\author{Wen-Bin Xu$^{1}$ and Shao-Feng Wu$^{1,2}$}
\affiliation{$^{1}$Department of Physics, Shanghai University, Shanghai 200444, China}
\affiliation{$^{2}$Center for Gravitation and Cosmology, Yangzhou University, Yangzhou 225009, China}
\author{}

\emailAdd{xwb@shu.edu.cn}
\emailAdd{sfwu@shu.edu.cn}

\abstract{
Based on the AdS/CFT correspondence, we study how to reconstruct bulk spacetime metrics by various quantum information measures on the boundary field theories, which include entanglement entropy, mutual information, entanglement of purification, and computational complexity according to the proposals of complexity=volume 2.0 and complexity=generalized volume. We present several reconstruction methods, all of which are free of UV divergence and most of which are driven by the derivatives of the measures with respect to the boundary scales. We illustrate that the exterior and interior of a black hole can be reconstructed using the measures of spatial entanglement and time-evolved complexity, respectively. We find that these measures always probe the spacetime in a local way: reconstructing the bulk metric in different radial positions requires the information at different boundary scales. We also show that the reconstruction method using complexity=volume 2.0 is the simplest and has a certain strong locality.
}


\begin{document}

\maketitle
\flushbottom

\section{Introduction}

Anti-de Sitter/conformal field theory (AdS/CFT) correspondence not only
provides a gravitational lens for strongly coupled quantum field theories
but also brings a wealth of insights for quantum gravity \cite%
{Maldacena:1997re,Gubser:1998bc,Witten:1998qj,Liu:2020rrn,Susskind:1998dq}.
The essential challenge in exploring this correspondence is to understand
how the boundary degrees of freedom of the CFT can be reorganized under
certain limits into the local gravitational physics in the bulk. This
ambitious program is widely termed as `bulk reconstruction'.

Bulk reconstruction is a non-trivial inverse problem that involves
holographic mapping from low to high dimensions. One of the most fascinating
branches in this program is the reconstruction of the metric of the
holographic spacetime\footnote{%
One can see other important branches in \cite%
{Harlow:2018fse,DeJonckheere:2017qkk,Hamilton:2006az,Kajuri2003}, especially
the bulk operator reconstruction.}. There are various methods regarding the
bulk metric reconstruction, which use different boundary physical
quantities, such as the source and expectation value of energy-momentum
tensor \cite{deHaro:2000vlm}, the singularities in the set of correlation
functions \cite{Hammersley:2006cp,Hubeny:2006yu}, the entanglement entropy
(EE) of boundary intervals \cite%
{Hammersley:2007ab,Hubeny:2012ry,Bilson:2008ab,Bilson:2010ff}, the
differential entropy that is a UV-finite combination of EE \cite%
{Balasubramanian:2013lsa,Myers:2014jia,Czech:2014ppa}, the divergence
structure of boundary $n$-point function \cite%
{Engelhardt:2016wgb,Engelhardt:2016crc}, the modular Hamiltonians of
boundary subregions \cite{Roy:2018ehv,Kabat:2018smf}, the Wilson loops
related to quark potential \cite{Hashimoto:2020mrx}, and four-point
correlators in an excited quantum state \cite{Caron-Huot2211}, among others.
It is worth noting that much of the work on metric reconstruction has been
driven by the idea that spacetime is built by quantum entanglement \cite%
{Takayanagi
T,Maldacena:2001kr,Ryu:2006bv,Swingle:2009bg,VanRaamsdonk:2010pw,Maldacena:2013xja}%
. However, it has been pointed out that `entanglement is not enough' to
encode the full spacetime \cite{Susskind:2014moa}. In order to understand
the interior of a black hole, it has been proposed that the quantum
computational complexity would be important. In fact, based on the
`complexity=volume' (CV) proposal \cite{Susskind:2014rva,Susskind:2014moa},
Hashimoto and Watanabe have successfully reconstructed the metric inside
black holes \cite{Hashimoto:2021umd}.

One common challenge in computing physical quantities in quantum field
theories is how to address the issue of UV divergence. In the AdS/CFT
correspondence, there are systematic methods to cancel the divergence based
on the UV/IR connection \cite{Susskind:1998dq}, which are well known as
holographic renormalization \cite{Skenderis0209,Papadimitriou2016}. However,
most of work on holographic renormalization is constrained to the standard
AdS/CFT, which requires the conformal symmetry, infinite coupling, and large
N limit. How to remove the divergence on gravity and field theory
consistently beyond these constraints remains an open question. In light of
this, ref. \cite{Jokela:2020auu} uses the derivative of EE with respect to
the spatial size of entangling region $l $ as data, which are insensitive to
the UV cutoff, unlike to EE itself. As a result, the reconstruction of
metric is free of UV divergence. Note that canceling off the UV divergence
has also been a guide to the correct formula for the differential entropy
\cite{Balasubramanian:2013lsa}.

On the other hand, deep learning (DL) algorithms have been utilized to
reconstruct the metric \cite%
{Hashimoto:2018ftp,Hashimoto:2018bnb,Tan:2019czc,Akutagawa:2020yeo,Hashimoto:2020jug,Yan:2020wcd,Hashimoto:2019bih,Hashimoto:2021ihd,Katsube:2022ofz,Hashimoto:2022eij,Li:2022zjc}%
. This not only unlocks the potential for building a data-driven holographic
model but also provides insights into holography using the language of
machine learning\footnote{%
This program has been referred as the `AdS/DL' correspondence. Other related
work that extracts the spacetime metric by machine learning but does not
rely on AdS/CFT can be found in \cite%
{You:2017guh,Hu:2019nea,Han:2019wue,Lam:2021ugb}.}. In ref. \cite%
{Yan:2020wcd}, an interesting difference was observed between the analytical
reconstruction method using the $l$-dependent EE as data \cite{Bilson:2010ff}
and the DL method using the frequency-dependent shear viscosity: the former
`locally' probes the bulk spacetime, while the latter is `non-local'. Here
the non-local aspect is manifested through the excellent generalization
capability of the deep neural network, which enable a narrow frequency to
generate a wide frequency shear viscosity, using the learned complete metric
as a hidden structure. Conversely, the local aspect indicates that the
deeper spacetime is probed through the EE with a larger entangling region.

In this paper, we will explore the reconstruction of bulk metric using
various quantum information measures on the boundary field theories. In
addition to revisiting EE, we will study mutual information (MI) \cite%
{Wolf:2007tdq} and entanglement of purification (EoP) \cite{Terhal2002},
both of which are closely related to EE, as well as two candidate measures
of complexity that differ from the CV proposal \cite%
{Couch:2016exn,Belin:2021bga}. For most of them, we will use their
derivatives with respect to boundary scales as data. Moreover, we will
compare how these measures encode the bulk metric. In particular, we will
examine whether they always probe the spacetime in a local way.

\section{Exterior of black holes}

\label{sec:two}

\subsection{Entanglement Entropy}

\label{subsec:two}

As a warm-up, we review the analytical reconstruction method using EE which
was proposed by Bilson \cite{Bilson:2010ff}. The relevant holographic
dictionary is the Ryu-Takayanagi formula \cite{Ryu:2006bv}, by which the EE
of the entangling region $A$ on the boundary can be calculated in the bulk
spacetime:%
\begin{equation}
S=\frac{Area(\gamma _{A})}{4G_{N}}.  \label{eq:sec 2 SS}
\end{equation}%
Here $G_{N}$ is Newton's constant and $\gamma _{A}$ is the minimal surface
which extends into the bulk and shares the boundary with $\partial A$.
Afterwards, we will set $G_{N}=1$ for convenience.

\subsubsection{Holographic calculation}

\label{subsec:Caculation of the area}

\begin{figure}[tbp]
\centering
\includegraphics[height=5.5cm]{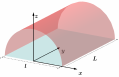}
\caption{A minimal surface (red) in the bulk and a strip entangling region
(blue) on the boundary. EE is given by the area of the red surface.}
\label{fig:strip3D}
\end{figure}

We will study a $d+1$-dimensional asymptotically AdS spacetime. As a proof
of principle, we focus on a highly symmetric metric ansatz%
\begin{equation}
ds^{2}=\frac{1}{z^{2}}\left( -f(z)dt^{2}+\frac{1}{f(z)}dz^{2}+d\vec{x}%
^{2}\right) .\,  \label{eq:sec 3 ansatz}
\end{equation}%
We set the planar horizon and boundary located at $z=1$ and $z=0$,
respectively. On the boundary, we are interested in a spatial strip
entangling region with a width $l$ in one direction and an infinite length $%
L $ in every other direction. The entangling region and minimal surface are
depicted in figure \ref{fig:strip3D}.
The area of the minimal surface is a function of $l$, which can be written
as
\begin{equation}
A(l)=2L^{d-2}\int_{0}^{\frac{l}{2}}\frac{1}{z^{d-1}}\sqrt{1+\frac{z^{\prime
2}}{f(z)}}dx,  \label{eq:sec 2 Area}
\end{equation}%
where $z^{\prime }=dz/dx$. The profile of the minimal surface can be
obtained as follows. Treating $A(l)$ as an action and $z(x)$ as a
generalized coordinate, one can read the Lagrangian
\begin{equation}
\mathcal{L}(z,z^{\prime })=\frac{1}{z^{d-1}}\sqrt{1+\frac{z^{\prime 2}}{f(z)}%
}.  \label{eq:sec 2 L}
\end{equation}%
It leads to the Hamiltonian
\begin{equation}
\mathcal{H}(z,z^{\prime })=z^{\prime }\frac{d\mathcal{L}}{dz^{\prime }}-%
\mathcal{L}=-\frac{z^{1-d}}{\sqrt{1+\frac{z^{\prime 2}}{f(z)}}}.\,
\label{eq:sec 2 H}
\end{equation}%
Since $\mathcal{L}$ does not depend on the variable $x$ explicitly, one can
set $\mathcal{H}$ as a constant%
\begin{equation}
\mathcal{H}=-z_{\ast }^{1-d},
\end{equation}%
where $z_{\ast }$ is given by $\left. z^{\prime }\right\vert _{z=z_{\ast
}}=0 $. This yields%
\begin{equation}
\frac{dz}{dx}=\pm \frac{\sqrt{f(z)(z_{\ast }^{2d-2}-z^{2d-2})}}{z^{d-1}}.
\label{eq:sec 2 dzdx}
\end{equation}%
Equation \eqref{eq:sec 2 dzdx} with the boundary conditions $z(\pm l/2)=0$
determines the profile function $z(x)$ for the minimal surface.

From \eqref{eq:sec 2 SS}, \eqref{eq:sec 2 Area} and \eqref{eq:sec 2 dzdx},
we have
\begin{eqnarray}
S(z_{\ast }) &=&\frac{L^{d-2}}{2}\int_{\delta }^{z_{\ast }}\frac{z_{\ast
}^{d-1}}{z^{d-1}\sqrt{f(z)}\sqrt{z_{\ast }^{2d-2}-z^{2d-2}}}dz,
\label{eq:sec 2 Sf} \\
\,l(z_{\ast }) &=&2\int_{0}^{z_{\ast }}\frac{z^{d-1}}{\sqrt{f(z)}\sqrt{%
z_{\ast }^{2d-2}-z^{2d-2}}}dz,  \label{eq:sec 2 lf}
\end{eqnarray}%
where $\delta $ is the UV cutoff. Note that when $\delta \rightarrow 0$, EE
is divergent. In order to cancel $z_{\ast }$ in two equations above, we take
derivatives:
\begin{equation}
\frac{dS(z_{\ast })}{dz_{\ast }}=\frac{L^{d-2}}{2}\left[ \frac{1}{\sqrt{%
f(z_{\ast })}}\lim_{z\rightarrow z_{\ast }}\frac{1}{\sqrt{z_{\ast
}^{2d-2}-z^{2d-2}}}-\int_{0}^{z_{\ast }}\frac{(d-1)z^{d-1}z_{\ast }^{d-2}dz}{%
(z_{\ast }^{2d-2}-z^{2d-2})^{\frac{3}{2}}\sqrt{f(z)}}\right] ,
\label{eq:sec 2 dSdz*}
\end{equation}

\begin{equation}
\frac{dl(z_{\ast })}{dz_{\ast }}=2z_{\ast }^{d-1}\left[ \frac{1}{\sqrt{%
f(z_{\ast })}}\lim_{z\rightarrow z_{\ast }}\frac{1}{\sqrt{z_{\ast
}^{2d-2}-z^{2d-2}}}-\int_{0}^{z_{\ast }}\frac{(d-1)z^{d-1}z_{\ast }^{d-2}dz}{%
(z_{\ast }^{2d-2}-z^{2d-2})^{\frac{3}{2}}\sqrt{f(z)}}\right] .
\label{eq:sec 2 dldz*}
\end{equation}%
Using the chain rule
\begin{equation}
\frac{dS(l)}{dl}=\frac{dz_{\ast }}{dl}\frac{dS(z_{\ast })}{dz_{\ast }},
\label{eq:sec 2 chain}
\end{equation}%
one can find
\begin{equation}
\frac{dS(l)}{dl}=\frac{L^{d-2}}{4z_{\ast }^{d-1}}.  \label{eq:sec 2 dSdl}
\end{equation}%
This equation builds a relation between $l$ and $z_{\ast }$.


\subsubsection{Reconstruction method}

\label{subsec: 2 Reconstruction method}

The key step of Bilson's method is to solve an integral equation
analytically. Referring to the handbook of integral equations \cite{Poly},
one can read:
\begin{align}
f(x)& =\int_{a}^{x}\frac{y(t)dt}{\sqrt{g(x)-g(t)}},\,g^{\prime }(x)>0  \notag
\\
\text{\textrm{Solution}}& \text{\textrm{: }}\,y(x)=\frac{1}{\pi }\frac{d}{dx}%
\int_{a}^{x}dt\frac{f(t)g^{\prime }(t)}{\sqrt{g(x)-g(t)}}.  \label{sol}
\end{align}%
Applying eq. (\ref{sol}) to eq. (\ref{eq:sec 2 Sf}), one can find%
\begin{equation}
\frac{1}{\sqrt{f(z)}}=\frac{4\left( d-1\right) z^{d-1}}{\pi L^{d-2}}\frac{d}{%
dz}\int_{\delta }^{z}\frac{S(z_{\ast })z_{\ast }^{d-2}}{\sqrt{%
z^{2d-2}-z_{\ast }^{2d-2}}}dz_{\ast }.  \label{Bilson}
\end{equation}%
This is the reconstruction formula provided by Bilson \cite{Bilson:2010ff}.
Here $S(z_{\ast })$ is generated by the data $S(l)$ and $l$ is related to $%
z_{\ast }$ through eq. (\ref{eq:sec 2 dSdl}).

The analytic formula (\ref{Bilson}) is simple. However, the field theory
data it inputs is $S(l)$, which is sensitive to the UV cutoff. Following
ref. \cite{Jokela:2020auu}, we will use the derivative $S^{\prime }(l)$ as
the data, which is insensitive to the UV cutoff. We apply eq. (\ref{sol}) to
eq. (\ref{eq:sec 2 lf}) instead of eq. (\ref{eq:sec 2 Sf}), which yields a
slightly different reconstruction formula\footnote{%
Note that the exact formula \eqref{eq:sec 2 f(z)} has not appeared before.
In \cite{Jokela:2020auu}, the metric is expanded using some basis functions
and their coefficients are related to $l(z_{\ast })$. The advantage of this
prescription is to increase the efficiency of numerical computation.}

\begin{equation}
\frac{1}{\sqrt{f(z)}}=\frac{d-1}{\pi z^{d-1}}\frac{d}{dz}\int_{0}^{z}\frac{%
l(z_{\ast })z_{\ast }^{2d-3}}{\sqrt{z^{2d-2}-z_{\ast }^{2d-2}}}dz_{\ast }.
\label{eq:sec 2 f(z)}
\end{equation}


\subsubsection{Example}

\label{subsec:checking}

\begin{figure}[tbp]
\centering
\includegraphics[scale=0.5]{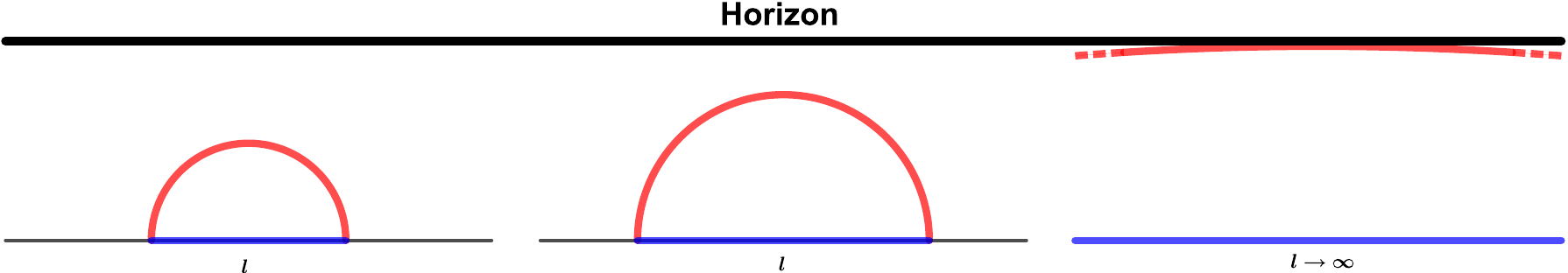}
\caption{The geodesic probes the deeper spacetime as $l$ increases. As $l$
tends to infinity, the geodesic will approach the horizon.}
\label{fig:2DEE1}
\end{figure}

We can check the validity of the reconstruction formula \eqref{eq:sec 2 f(z)}
with a simple example. We choose the AdS black hole with $d=2$ and the
function $f(z)=1-z^{2}$ as our target metric. Referring to the result of
\cite{Fischler:2012uv}, the EE of a strip of width $l$ is given by
\begin{equation}
S(l)=\frac{1}{2}\log \left[ \frac{2}{\delta }\sinh \left( \frac{l}{2}\right) %
\right] .  \label{eq:sec 2 sfz}
\end{equation}%
In our task of bulk reconstruction, the metric is unknown but we are
supposed to know%
\begin{equation}
S^{\prime }(l)=\frac{1}{4}\coth (\frac{l}{2}).
\end{equation}%
Combining the data with \eqref{eq:sec 2 dSdl}, we find%
\begin{equation}
l(z_{\ast })=2\,\text{arctanh}(z_{\ast }).  \label{eq:sec 2 l(z*)}
\end{equation}%
This formula indicates that when $l$ increases, the top of the geodesic
probes the deeper spacetime near the horizon. We visualize the probing way
in figure \ref{fig:2DEE1}. At last, substituting \eqref{eq:sec 2 l(z*)} into %
\eqref{eq:sec 2 f(z)} with $d=2$, we recover our target metric analytically
\begin{equation}
f(z)=1-z^{2}.  \label{eq:sec 2 fz}
\end{equation}

\subsection{Mutual Information}

\label{subsec:MI}

MI is a closely related to EE. It measures the total (both classical and
quantum) correlations between two spatial subregions and acts as an upper
bound of the connected correlation functions in those regions \cite%
{Wolf:2007tdq,Swingle:2010jz}.\ The definition of MI between two disjoint
subsystems $A$ and $B$ is given by
\begin{equation}
I(A,B)=S_{A}+S_{B}-S_{A\cup B},  \label{eq: I}
\end{equation}%
where $S_{A}$, $S_{B}$ and $S_{A\cup B}$ denote the EE of the region $A,B$
and ${A\cup B}$ respectively with the rest of the system \cite%
{Fischler:2012uv}. Note that EE is a divergent quantity, but MI is finite
since three divergent parts in eq. (\ref{eq: I}) that depend on the UV
cutoff cancel out.

\subsubsection{Holographic calculation}

\label{subsec: Reconstruction formula of mutual information}

\begin{figure}[tbp]
\centering
\includegraphics[height=5.5cm]{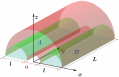}
\caption{The minimal surfaces (red and green) and two disjoint strip
entangling regions (blue) in the symmetric configuration. The nonzero MI is
given by the difference between the area of green surfaces and red surfaces.}
\label{fig:stripAB}
\end{figure}

Consider two disjoint subsystems with strip regions $A$ and $B$, each with
width $l$ and length $L\rightarrow \infty $. The interval between them is $a$%
. For the sake of simplicity, we focus on the symmetric configuration, see
figure \ref{fig:stripAB}. In this case, $S_{A\cup B}=\min \left[
2S(l),S(2l+a)+S(a)\right] $. Depending on the ratio $a/l$, the MI can be
written as
\begin{equation}
I(l)=%
\begin{cases}
2S(l)-S(2l+a)-S(a)\,\,\,\,\text{for small ratio $a/l$}, \\
\,\,\,\,\,\,\,\,\,\,\,\,0\,\,\,\,\,\,\,\,\,\,\,\,\,\,\,\,\,\,\,\,\,\,\,\,\,%
\,\,\,\,\,\,\,\,\,\text{for large ratio $a/l$}.%
\end{cases}
\label{eq: I(l)}
\end{equation}%
For convenience, we will fix $a$ and change $l$. This induces a critical
width $l_{a}$ below which the MI is zero. Note that one cannot reconstruct
anything from the region where MI is zero. In fact, we have not found\ a way
to reconstruct the metric through the MI alone.

\subsubsection{Reconstruction method}

\label{subsec: 3 Reconstruction method}

Here we will propose a method which reconstructs the metric by jointly using
EE and MI. Suppose that the field theory data we have are the derivatives of
MI and EE with different ranges of $l$,\footnote{An interesting question is whether it is possible in reality to know only MI but not EE. MI has some nice properties that EE does not have, such as being finite and non-negative. In view of this, we suspect that in certain situations, it is advantageous to measure the MI directly rather than the EE. At that time, one can know the MI alone.}%
\begin{equation}
I^{\prime }(l),\;l>l_{a}\text{ and }S^{\prime }(l),\;l\leq 2l_{a}+a,
\label{IS}
\end{equation}%
where $a$ (then $l_{a}$) is fixed. A simple but important observation is
that one can combine eq. (\ref{eq: I(l)}) and the data (\ref{IS}) to
iteratively generate $S^{\prime }(l)$ with any $l$. The iterative equation
can be written as%
\begin{equation}
S^{\prime }\left( Q\right) =\left\{
\begin{array}{c}
\;\;S^{\prime }(Q),\;\;\;\;\;\;\;\;\;\;\;\;\;\;\ Q\leq Q_{a}\;\;\;\;\;\; \\
S^{\prime }\left( \frac{Q-a}{2}\right) -\frac{1}{2}I^{\prime }\left( \frac{%
Q-a}{2}\right) ,\ Q_{a}<Q\leq 2Q_{a}+a%
\end{array}%
\right.
\end{equation}%
where $Q_{a}=2l_{a}+a$. After the $n$-th iteration, one can generate the
data $S^{\prime }\left( Q\right) $ with $Q\leq 2nQ_{a}+(2n-1)a$. Then the
metric can be reconstructed using (\ref{eq:sec 2 dSdl}) and (\ref{eq:sec 2
f(z)}).


\subsubsection{Example}

\label{subsec: 3 Checking}

\begin{figure}[tbp]
\centering
\includegraphics[scale=0.5]{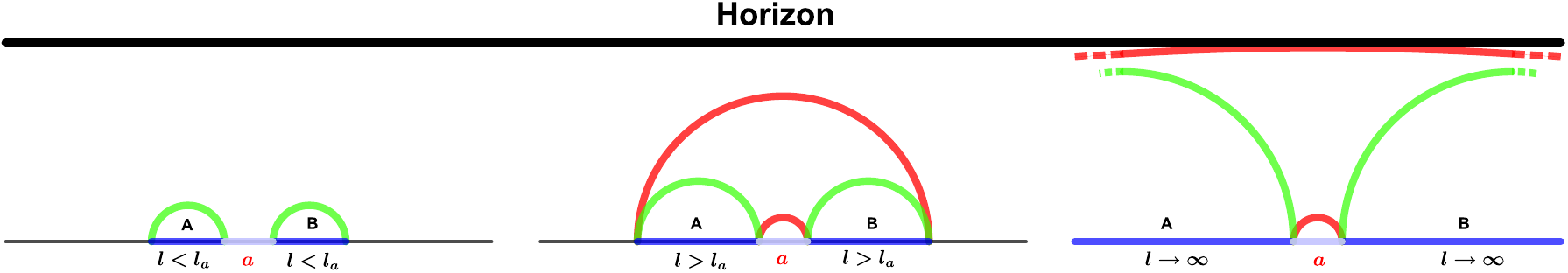}
\caption{The geodesics relevant to MI vary as $l$ increases. As $l$ tends to
infinity, the external red geodesic and two green geodesics approach the
horizon in turn.}
\label{fig:2DMI}
\end{figure}

\begin{figure}[tbp]
\centering
~~\includegraphics[scale=0.45]{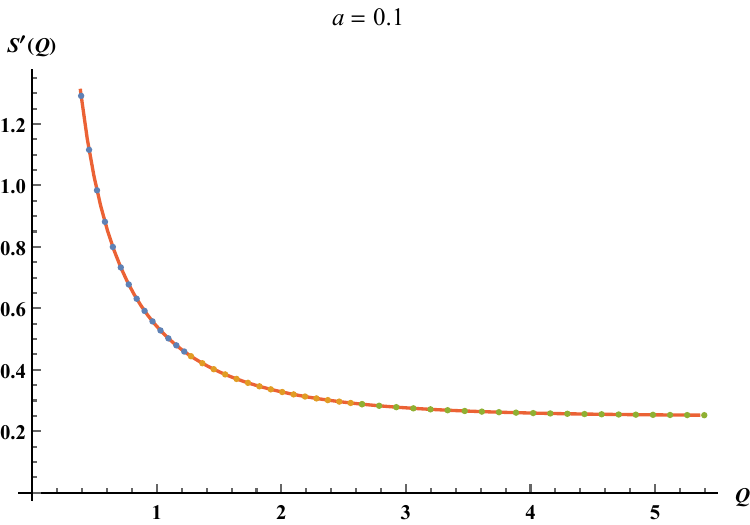} \hspace{0.2in} %
\includegraphics[scale=0.45]{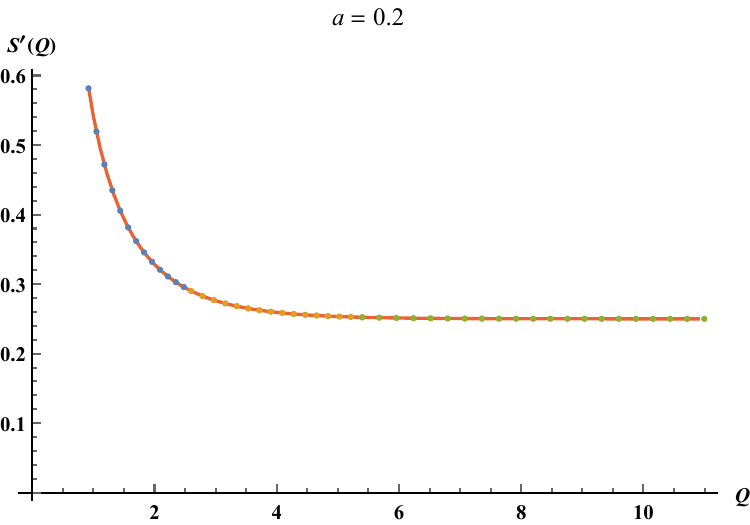}
\caption{Generate $S^{\prime }(Q)$ iteratively. Here we set $a=0.1$ (left)
and $a=0.2$ (right). The solid red line is plotted using the target
analytical function. The dotted lines denote the generated function. By
three iterations (blue, yellow, green), the domain of $S^{\prime }(Q)$
expands accordingly.}
\label{fig:MISQ}
\end{figure}

\begin{figure}[tbp]
\centering
\includegraphics[scale=0.45]{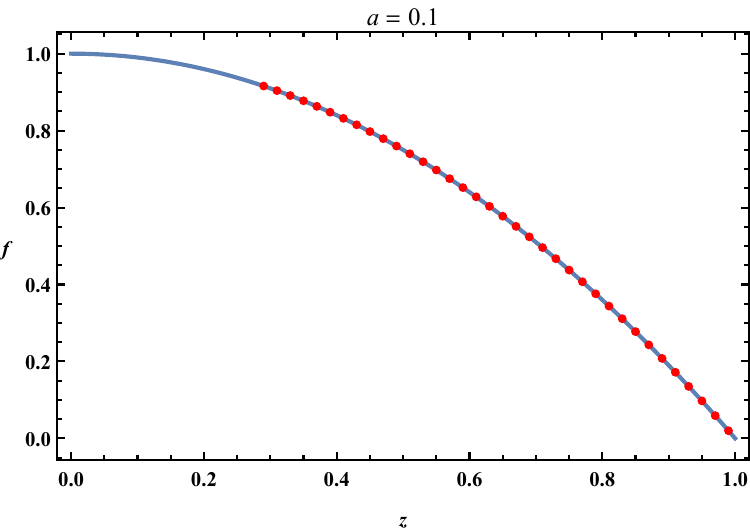} \hspace{0.2in} %
\includegraphics[scale=0.45]{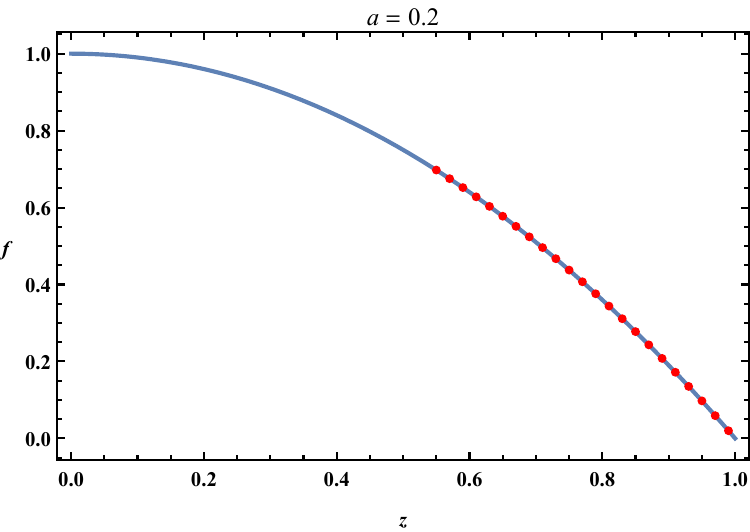}
\caption{The reconstructed metric over $Z_{\ast a}$ (red dotted line) using
MI and EE. The target metric (blue solid line) is $f=1-z^{2}$. At left and
right panels, we set $a=0.1$ and $0.2$ respectively.}
\label{fig:MIa}
\end{figure}

Let's exhibit our reconstruction method with an example. We still consider $%
d=2$ and $f(z)=1-z^{2}$. Using eq. (\ref{eq:sec 2 sfz}) and eq. (\ref{eq:
I(l)}), we prepare our data as%
\begin{equation}
I^{\prime }(l)=\frac{1}{2}\left[ \coth (\frac{l}{2})-\coth (\frac{a}{2}+l)%
\right] ,\ l>l_{a}  \label{eq: Il}
\end{equation}%
and%
\begin{equation}
S^{\prime }(l)=\frac{1}{4}\coth (\frac{l}{2}),\ l\leq 2l_{a}+a.  \label{Sla}
\end{equation}%
Here we can specify the critical width%
\begin{equation*}
l_{a}=\text{arccosh}\left[ \frac{\sqrt{2}\cosh \left( \frac{a}{2}\right)
+5\cosh \left( a\right) -\sqrt{2}\cosh \left( \frac{3a}{2}\right) -7}{8\cosh
\left( a\right) -10}\right] .
\end{equation*}%
The associated region where MI is non-zero can be found in figure \ref%
{fig:MI}. Note that $Q_{a}=2l_{a}+a$ corresponds to a critical point $%
Z_{\ast a}$ in the bulk%
\begin{equation}
Z_{\ast a}=\tanh \left[ 2\,\text{arccoth}\left[ (3-2\sqrt{2})\coth \left(
\frac{a}{4}\right) \right] \right] ,
\end{equation}%
upon which is the bulk region that cannot be probed by the derivative of EE (%
\ref{Sla}) alone. In figure \ref{fig:2DMI}, we fix $a$ and plot the change
of the relevant geodesics as $l$ increases. In figure \ref{fig:MISQ}, we
iteratively generate $S^{\prime }(Q)$ beyond $Q_{a}$. In figure \ref{fig:MIa}%
, we reconstruct the metric.


\begin{figure}[tbp]
\centering
~~~~\includegraphics[scale=0.7]{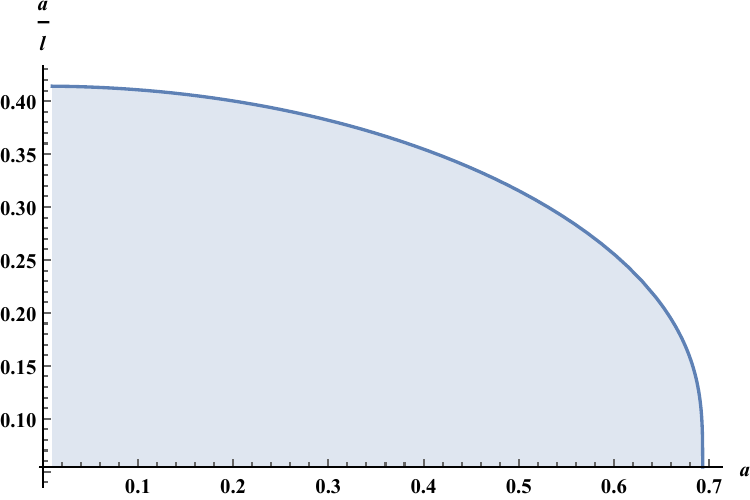}
\caption{The MI is non-zero only in the blue shaded region.}
\label{fig:MI}
\end{figure}

\subsection{Entanglement of Purification}

\label{subsec:EoP}

EoP is defined by the minimum EE for all possible purification of the mixed
state \cite{Terhal2002}\footnote{%
Note that a slight generalization of EoP has been used to study the
reconstruction of spacelike curves within the entanglement wedge \cite%
{Espindola:2018ozt}.}. It is closely related to MI but is an independent correlation measure \cite{Bagchi2015}. The regularization of EoP has an operational meaning in terms of Einstein-Podolsky-Rosen pairs \cite{Terhal2002}, and it is finite in the CFT \cite{Caputa1812}. In gravity side it was proposed to be related to the
entanglement wedge cross section \cite{Tamaoka:2018ned,Takayanagi:2017knl}%
\begin{equation}
E_{W}=\min_{\Sigma _{AB}}\left( \frac{Area(\Sigma _{AB})}{4}\right) .
\label{eq:sec 2 EW}
\end{equation}%
Here $A$ and $B$ represent two disjoint subregions on the boundary. The
entanglement wedge is a bulk region surrounded by $AB\equiv A\cup B$ and the
minimal surface $\Gamma _{AB}$ anchored on $AB$. $\Sigma _{AB}$ is regarded
as a cross section of the entanglement wedge, which divides $\Gamma _{AB}$
into two parts. $E_{W}$ can be obtained by minimizing the area of $\Sigma
_{AB}$ over all possible choices of the division.

\begin{figure}[tbp]
\centering
\includegraphics[height=5.5cm]{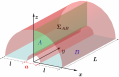}
\caption{The entanglement wedge with non-zero EoP is a connected region
wrapped by the red and blue surfaces. The green surface is the minimum cross
section. EoP is determined by its area.}
\label{fig:EoP}
\end{figure}

In figure \ref{fig:EoP}, we plot the geometry for non-zero EoP. We still
focus on the symmetric configuration for simplicity. One can find that the
EoP is zero when the MI is zero. This is because the entanglement wedge is
disconnected and lacks a cross section.

\subsubsection{Holographic calculation}

\label{sec:4}

Consider the situation with $E_{W}\neq 0$. The area of $\Sigma _{AB}$ is
\begin{equation}
A_{\Sigma }=L^{d-2}\int_{z_{\ast }(a)}^{Z_{\ast }(Q)}\frac{dz}{\sqrt{f(z)}%
z^{d-1}},\,Q>Q_{a},  \label{eq:sec 4 A}
\end{equation}%
where $z_{\ast }(a)$ and $Z_{\ast }(Q)$ are the turning points of the
minimal surfaces with strips of width $a$ and $Q$, respectively. Using %
\eqref{eq:sec 2 EW}, we have
\begin{equation}
E_{W}(Z_{\ast })=\frac{L^{d-2}}{4}\int_{z_{\ast }}^{Z_{\ast }}\frac{dz}{%
\sqrt{f(z)}z^{d-1}},\,Z_{\ast }>Z_{\ast a}.  \label{eq:sec 4 E}
\end{equation}%
Obviously, it is finite as $z_{\ast }>0$. Take the derivative of eq. (\ref%
{eq:sec 4 E}), yielding
\begin{equation}
E_{W}^{\prime }(Z_{\ast })=\frac{L^{d-2}}{4}\frac{1}{\sqrt{f(Z_{\ast })}%
Z_{\ast }^{d-1}},\,Z_{\ast }>Z_{\ast a}.  \label{E'}
\end{equation}%
Because this formula is structurally different from eq.
\eqref{eq:sec 2
dldz*}, we cannot use the simple chain rule as before to find the
metric-independent mapping $Q\mapsto Z_{\ast }$ from the change rate of the
EoP with respect to $Q$.


\subsubsection{Reconstruction method}

\label{sec:4 Joint} We will propose a numerical method, which include four
steps.

1) Discretization

Let's discrete the metric function as $f[z[i]]$, where $i$ denotes an
integer in $[0,n]$, $z[i]=i/n$, and $n$ is a large integer. Suppose that the
metric $f(z)$ with $z\leq Z_{\ast }$ is given, where $Z_{\ast }$ belongs to $%
z[j]\leq Z_{\ast }<z[j+1]$ with the integer $j<n$. Build the dataset $%
\{z[i],f[z[i]]\}$ for any $i\in \lbrack 0,j]$ and name it as $data_{0}$.

2) Interpolation

Append a data point $\{z[j+1],f_{1}\}$ into $data_{0}$. The value of $f_{1}$
will be specified later and the new dataset is named as $data_{1}$.
Interpolate $data_{1}$ and generate the test metric function $\bar{f}(z)$
which is continuous.

3) Integration

Calculate the numerical integration%
\begin{eqnarray}
Q(Z_{\ast }) &=&2\int_{0}^{Z_{\ast }}\frac{z^{d-1}}{\sqrt{\bar{f}(z)}\sqrt{%
Z_{\ast }^{2d-2}-z^{2d-2}}}dz,  \label{QZS} \\
\bar{E}_{W}(Z_{\ast }) &=&\frac{L^{d-2}}{4}\int_{z_{\ast }}^{Z_{\ast }}\frac{%
dz}{\sqrt{\bar{f}(z)}z^{d-1}},\,Z_{\ast }>Z_{\ast a}.  \label{EWZS}
\end{eqnarray}

4) Minimization

Define a loss function%
\begin{equation}
\mathrm{loss}(f_{1})=\left\vert \bar{E}_{W}(Z_{\ast })-E_{W}(Q(Z_{\ast
}))\right\vert ^{2},  \label{loss}
\end{equation}%
where $E_{W}(Q)$ is the given data. Substitute eq. (\ref{QZS}) and eq. (\ref%
{EWZS}) into eq. (\ref{loss}). Minimize the loss function and get the
optimal $f_{1}$. After that, set $f[z[j+1]]=f_{1}$, $j=j+1$, and $%
data_{0}=data_{1}$.

Finally, iterate the steps 2), 3) and 4). Then the metric with $z>Z_{\ast }$
can be reconstructed. Note that this method is characterized by the use of a
continuous interpolation function to generate the test metric and thereby we
would like to refer it as the interpolation-generated method.


\subsubsection{Example}

\label{subsec:2.3.3}

\begin{figure}[tbp]
\centering
\!\!\!\!\!\includegraphics[scale=0.6]{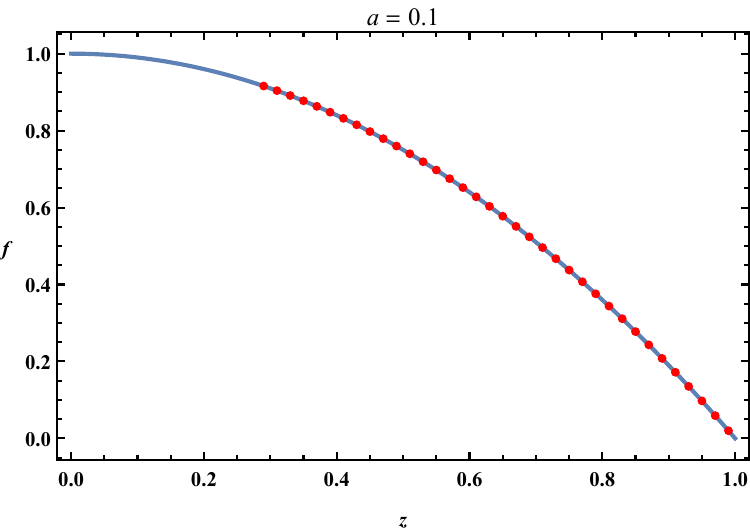}
\caption{The reconstructed metric over $Z_{\ast a}$ (red dotted line) using
the EoP, provided that the metric below $Z_{\ast a}$ is given. The target
metric (blue solid line) is $f=1-z^{2}$. Here we set $a=0.1$ and $n=100$.}
\label{fig:EoPf}
\end{figure}

For $d=2$ and $f(z)=1-z^{2}$, the EoP is \cite{Yang1810}%
\begin{equation}
E_{W}(Q)=\frac{1}{4}\log \left[ \coth \left( \frac{a}{4}\right) \tanh \left(
\frac{Q}{4}\right) \right] ,\ Q>Q_{a},  \label{eq:sec 3 EWD}
\end{equation}%
which provides us the main data. In addition, we need the metric below $%
Z_{\ast a}$, which can be given by hand or generated by the EE with $Q\leq
Q_{a}$. The rest of the metric can be reconstructed using EoP. In figure \ref%
{fig:2DEOP}, we plot the change of the entanglement wedge and minimum cross
section as $l$ increases. In figure \ref{fig:EoPf}, we show the result of
reconstruction\footnote{%
Using the canned tools in Mathematica (Interpolation, NIntegrate and
FindMinimum) with the option WorkingPrecision$\rightarrow $30, the
reconstruction task can be completed in a few minutes for an average laptop,
with a maximum relative error (occurring near the horizon) of $10^{-12}$.},
where we have specified $a=0.1$ and $n=100$. It is consistent with the left
panel of figure \ref{fig:MIa} as expected.

\begin{figure}[tbp]
\centering
\includegraphics[scale=0.5]{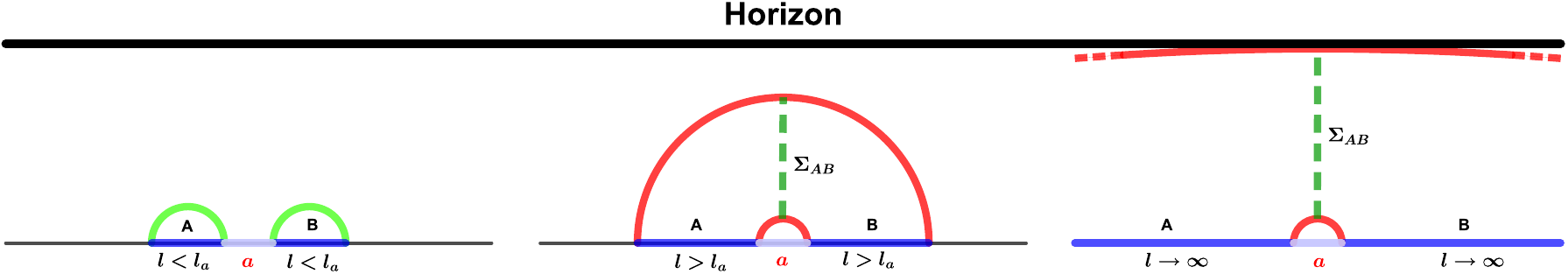}
\caption{The entanglement wedge and its minimum cross section vary as $l$
increases. As $l$ tends to infinity, the top of the minimum cross section
approaches the horizon.}
\label{fig:2DEOP}
\end{figure}

\section{Interior of black holes}

\label{sec:2}

Quantum computational complexity is relevant to the number of unitary
operators which converts one quantum state to another \cite{Watrous0804}. In
\cite{Hashimoto:2021umd}, it is found that the interior of a black hole can
be reconstructed from the complexity, which is defined by `complexity =
volume'\ (CV) proposal \cite{Susskind:2014rva,Susskind:2014moa}. There are
other proposals on the holographic complexity, such as `complexity =
action'\ (CA) \cite{Brown:2015bva} and `complexity = volume 2.0'\ (CV2.0)
\cite{Couch:2016exn}. Recently, a new infinite class of gravitational
observables has been proposed to be dual to the complexity \cite%
{Belin:2021bga}, which can be referred as `complexity = generalized volume'\
(CGV). In this section, we will study how to reconstruct the metric in terms
of CV2.0 and CGV\footnote{%
We do not know how to reconstruct using the CA proposal. Note that we cannot
specify the action that serves as a metric functional. Otherwise the
metric can be obtained through the variation. We suspect that such a
reconstruction method is highly non-trivial, if existed.}.

\subsection{Complexity = Volume 2.0}

The `complexity 2.0'\ in the gravity side can be expressed as \cite%
{Couch:2016exn}
\begin{equation}
\mathcal{C}_{2.0}(t)=\frac{P}{\hbar }V(t).  \label{eq:sec 2 S}
\end{equation}%
Here, $P$ is the pressure which is related to the cosmological constant $%
P=-\Lambda /8\pi $ and $V(t)$ is the spacetime volume of the Wheeler-DeWitt
(WDW) patch defined in a two-sided eternal AdS black holes. Note that the
WDW patch is the domain of dependence of any Cauchy surface in the bulk
which is anchored at the time slices on the boundary \cite%
{Brown:2015bva,Carmi:2016wjl}.


\subsubsection{Holographic calculation}

\label{subsec: 1}

\begin{figure}[tbp]
\centering
\!\!\!\!\!\!\!\!\!\!\!\!\!\!\!\!\!\!\!\!\!\!%
\includegraphics[scale=0.75]{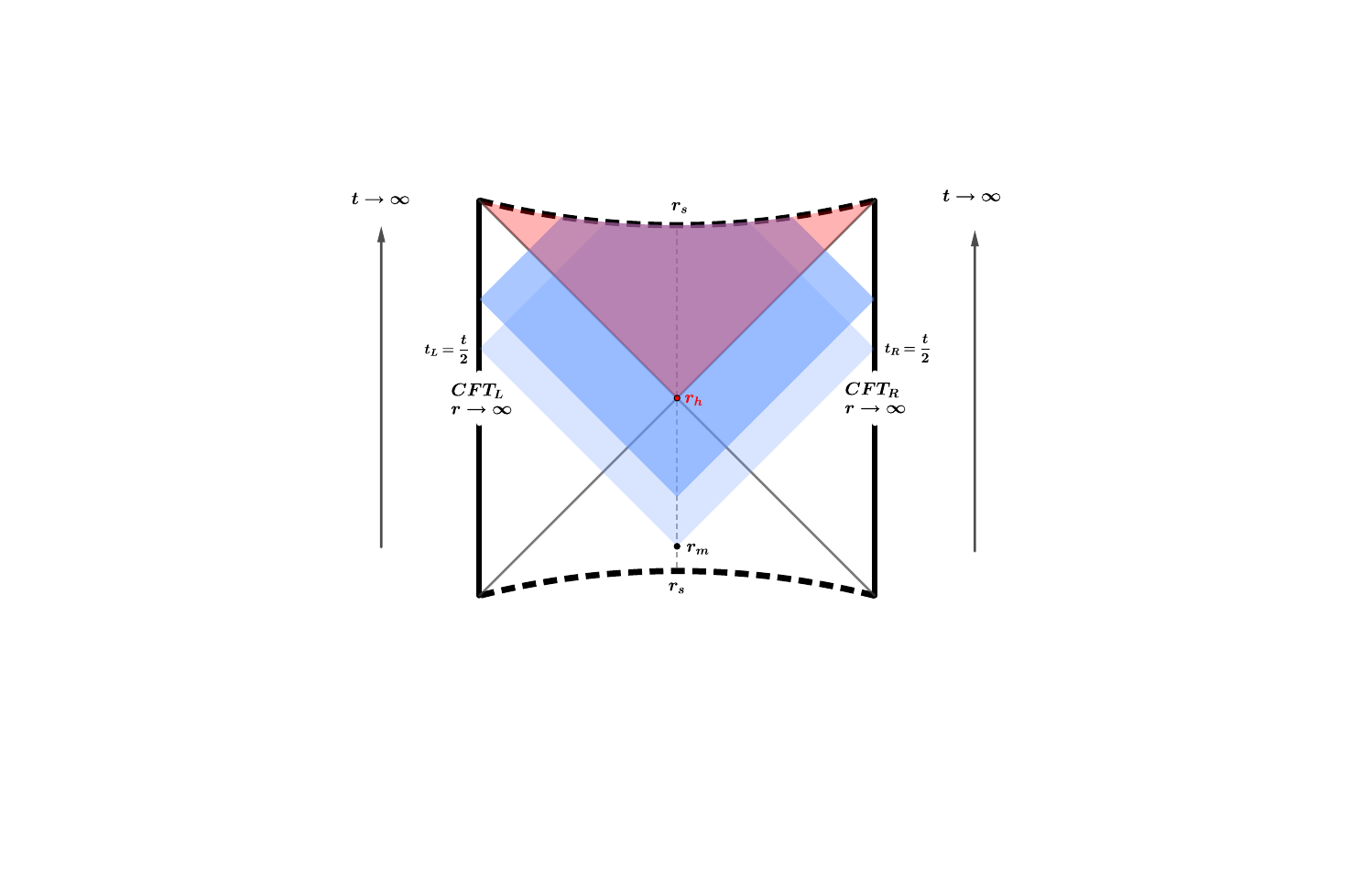} \hspace{0.5in} %
\includegraphics[scale=0.4]{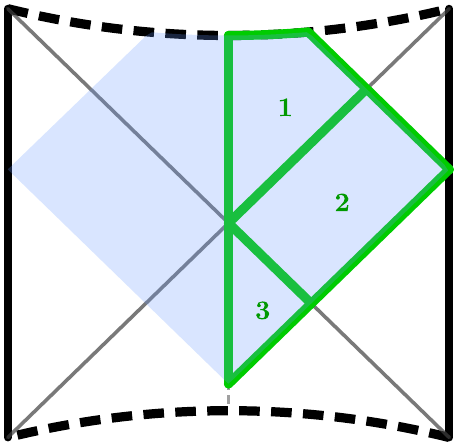}
\caption{Left: The time evolution of the WDW patch in a two-sided AdS black
hole. It is anchored at two boundary time slices $t_{L}=t/2$ and $t_{R}=t/2$%
. Two lower null sheets intersect at point $r_{m}$ and two upper null sheets
end at the singularity $r_{s}$. As time grows, the WDW patch moves upward,
through three shaded areas. The intersection $r_{m}$ will eventually
approach the horizon $r_{h}$. Right: Three typical blocks of the WDW patch
are labeled for the calculation of its volume.}
\label{fig:PenCV2}
\end{figure}

We first review the calculation of the spacetime volume $V(t)$ under the
metric ansatz%
\begin{equation}
ds^{2}=-f(r)dt^{2}+\frac{1}{f(r)}dr^{2}+r^{2}d\vec{x}^{2}.\,
\end{equation}
We consider that the WDW patch moves forward in time in a symmetric way and
assume that its two upper null sheets always end at the singularity\footnote{%
This holds for a class of black holes which have the Penrose diagrams with
similar structures.}, see the left panel of figure \ref{fig:PenCV2}.
Using the tortoise coordinate%
\begin{equation*}
r^{\ast }(r)=\int \frac{dr}{f(r)},
\end{equation*}%
one can determine the critical time until which the complexity does not
increase \cite{Carmi:2017jqz}
\begin{equation}
t_{c}=2(r^{\ast }\left( \infty \right) -r^{\ast }(r_{s})),
\label{eq:subsec 1 cri}
\end{equation}%
where $r_{s}$ denotes the position of singularity. To calculate the
spacetime volume, we divide the WDW patch into several parts. Due to the
symmetric configuration, we just need to calculate the volumes of three
regions which are labeled in the right panel of figure \ref{fig:PenCV2}.
Using the integration formula of the spacetime volume
\begin{equation}
V=\int_{WDW}d^{d+1}x\sqrt{-g}=\Sigma _{d-1}\int dtdrr^{d-1}
\label{eq:subsec 1 spac}
\end{equation}%
where $\Sigma _{d-1}$ denotes the regulated volume of the $d-1$ spatial
boundary\ directions, we have
\begin{equation}
V_{1}=\Sigma _{d-1}\int_{r_{s}}^{r_{h}}drr^{d-1}(t_{R}+r^{\ast }(\infty
)-r^{\ast }(r)),  \label{subsec:1 1}
\end{equation}

\begin{equation}
V_{2}=2\Sigma _{d-1}\int_{r_{h}}^{r_{\mathrm{uv}}}drr^{d-1}(r^{\ast }(\infty
)-r^{\ast }(r)),  \label{subsec:1 2}
\end{equation}

\begin{equation}
V_{3}=\Sigma _{d-1}\int_{r_{m}}^{r_{h}}drr^{d-1}(-t_{R}+r^{\ast }(\infty
)-r^{\ast }(r)),  \label{subsec:1 3}
\end{equation}%
where $r_{m}$ is determined by
\begin{equation}
t-t_{c}=2(r^{\ast }(r_{s})-r^{\ast }(r_{m}))  \label{subsec:1 t}
\end{equation}%
and $r_{\mathrm{uv}}$ is the UV cutoff. As a result, one can see \cite%
{An:2018dbz}
\begin{equation}
\frac{dV}{dt}=\frac{\Sigma _{d-1}}{d}\left( r_{m}^{d}-r_{s}^{d}\right) .
\label{subsec:1 dV}
\end{equation}%
By setting $\delta t=t-t_{c}$ and $r_{s}=0$, we rewrite eq.
\eqref{subsec:1
dV} and eq. \eqref{subsec:1 t} as%
\begin{equation}
\mathcal{C}_{2.0}^{\prime }(\delta t)=\frac{P}{\hbar }\frac{\Sigma _{d-1}}{d}%
r_{m}^{d},  \label{subsec:1 ct}
\end{equation}%
\begin{equation}
\delta t=-2\int_{0}^{r_{m}}\frac{dr}{f(r)}.  \label{subsec:1 t2}
\end{equation}


\subsubsection{Reconstruction method}

Our data is $\mathcal{C}_{2.0}^{\prime }(\delta t)$ of the boundary field
theory. Inputting the data into eq. (\ref{subsec:1 ct}), we can calculate
the derivative $\delta t^{\prime }(r_{m})$. Taking the derivative on both
sides of eq. (\ref{subsec:1 t2}), we obtain a very simple formula to
reconstruct the metric inside the horizon%
\begin{equation}
{f(r_{m})}=-\frac{2}{\delta t^{\prime }(r_{m})}.  \label{subsec:1 kx4}
\end{equation}

In ref. \cite{Hashimoto:2021umd}, the reconstruction of the interior of
black holes requires information from the exterior. Here, we have shown that
this is not necessary as soon as we update the proposal from CV to CV2.0.


\subsubsection{Example}

\label{sss: eg}

Suppose $\,f(r)=r^{2}-1$ and $d=2$. Calculating eq. \eqref{subsec:1 t2} and
taking the inverse, we find
\begin{equation}
r_{m}=\tanh \left( \frac{\delta t}{2}\right) .  \label{subsec:1 kx1}
\end{equation}%
Inserting it into \eqref{subsec:1 ct}, we read the complexity growth
\begin{equation}
\mathcal{C}_{2.0}^{\prime }(\delta t)=\frac{P}{\hbar }\frac{\Sigma _{1}}{2}%
\tanh ^{2}\left( \frac{\delta t}{2}\right) .  \label{subsec:1 ctd}
\end{equation}%
With the data (\ref{subsec:1 ctd}) in hand, we solve eq. \eqref{subsec:1 ct}
to give\footnote{%
Interestingly, this boundary-bulk relation encoded in CV2.0 has the same
form as the one encoded in EE, see eq. (\ref{eq:sec 2 l(z*)}).}
\begin{equation}
\delta t(r_{m})=2\,\text{arctanh}(r_{m}).  \label{subsec:1 tr}
\end{equation}%
Using the reconstruction formula \eqref{subsec:1 kx4}, we obtain the metric
finally
\begin{equation}
f(r_{m})=r_{m}^{2}-1.  \label{subsec:1 fr}
\end{equation}%
%
%
%

\subsection{Complexity = Generalized Volume}

Recently, the CV proposal has been generalized in \cite{Belin:2021bga}. It
is presented that the holographic dual of quantum complexity may be
expressed as an integral of a scalar functional $F_{1}$, which depends on
the background metric $g_{\mu \nu }$ and the embedding $X^{\mu }(\sigma
^{a}) $ of the codimension-one hypersurface $\Sigma _{F_{2}}$. The
hypersurface $\Sigma _{F_{2}}$ is anchored on the time slice $\Sigma _{t}$
of the boundary and is determined by extremizing another functional $F_{2}$.
In general, $F_{1}$ is different from $F_{2}$. For the simple case $%
F_{1}=F_{2}$, the generalized volume can be written as%
\begin{equation}
\mathcal{C}_{gen}(t)=\max_{\partial \Sigma =\Sigma _{t}}\left[ \frac{1}{L}%
\int_{\Sigma (t)}d^{d}\sigma \sqrt{h}F_{1}\left( g_{\mu \nu },X^{\mu
}\right) \right] ,  \label{Cgen0}
\end{equation}%
where $L$ is the AdS radius and $h$ is the determinant of induced metric.
Here we will explore the reconstruction using eq. (\ref{Cgen0}), provided $%
F_{1}$ depends only on the curvature invariant%
\begin{equation*}
F_{1}=1+\mathcal{K}L^{4}C^{2},
\end{equation*}%
where $C^{2}\equiv C_{\mu \nu \rho \sigma }C^{\mu \nu \rho \sigma }$ denotes
the square of the Weyl tensor for the bulk spacetime and $\mathcal{K}$ is
the coupling constant \cite{Belin:2021bga}.

\subsubsection{Holographic calculation}

Let's write the line element in the Eddington-Finkelstein coordinates
\begin{equation}
ds^{2}=-f(r)dv^{2}+2dvdr+r^{2}d\vec{x}^{2},  \label{EF ansatz}
\end{equation}%
where $v=t+r^{\ast }(r)$. By describing the extremal hypersurface by the
parametric equations $v(\sigma )$ and $r(\sigma )$, the generalized volume
can be rewritten in a specific form \cite{Belin:2021bga}
\begin{equation}
\mathcal{C}_{gen}=\Sigma _{d-1}\int_{\Sigma }d\sigma \,r^{d-1}\sqrt{-f(r)%
\dot{v}^{2}+2\dot{v}\dot{r}}a(r).  \label{Cgen}
\end{equation}%
Here we have set $L=1$ and $a(r)$ is a factor about the curvature, given by
\begin{equation}
a(r)=1+\mathcal{K}\frac{d-2}{d}{\frac{{1}}{{r^{4}}}\left[ 2f(r)+r\left(
rf^{\prime \prime }(r)-2f^{\prime }(r)\right) \right] }^{2}.  \label{ar}
\end{equation}%
We can express the generalized volume $\mathcal{C}_{gen}$ and the boundary
time $t$ in terms of the metric $f(r)$ and the minimal radius $r_{min}$ of
the surface%
\begin{equation}
\mathcal{C}_{gen}=-2\Sigma _{d-1}\int_{r_{min}}^{r_{\mathrm{uv}}}\frac{U(r)}{%
f(r)\sqrt{U(r_{min})-U(r)}}dr,  \label{sCCgen}
\end{equation}%
\begin{equation}
t=-2\int_{r_{min}}^{\infty }\frac{\sqrt{U(r_{min})}}{f(r)\sqrt{%
U(r_{min})-U(r)}}dr,  \label{sttgen}
\end{equation}%
where $U(r)=-f(r)\,a^{2}(r)\,r^{2(d-1)}$ can be understood as an effective
potential of a classical particle \cite{Belin:2021bga}. In order to cancel $%
r_{min}$ in two equations above, we take derivatives:
\begin{equation}
\frac{d\mathcal{C}_{gen}}{dr_{min}}\!=\!2\Sigma _{d-1}\!\left\{
\!\int_{r_{min}}^{\infty }\frac{U(r)U^{\prime }(r_{min})}{%
2f(r)[U(r_{min})-U(r)]^{\frac{3}{2}}}\!+\!\frac{U(r_{min})}{\sqrt{f(r_{min})}%
}\lim_{r\rightarrow r_{min}}\frac{1}{\sqrt{U(r_{min})-U(r)}}\!\right\} ,
\label{dSdrm}
\end{equation}%
\begin{equation}
\frac{dt}{dr_{min}}\!=\!\frac{2}{\sqrt{U(r_{min})}}\left\{
\int_{r_{min}}^{\infty }\frac{U(r)U^{\prime }(r_{min})}{%
2f(r)[U(r_{min})-U(r)]^{\frac{3}{2}}}\!+\!\frac{U(r_{min})}{\sqrt{f(r_{min})}%
}\lim_{r\rightarrow r_{min}}\frac{1}{\sqrt{U(r_{min})-U(r)}}\right\} .
\label{dtdrm}
\end{equation}%
Using the chain rule, the time derivative of the generalized volume is given
by
\begin{equation}
\frac{d\mathcal{C}_{gen}}{dt}=\Sigma _{d-1}\sqrt{U(r_{min})}.  \label{dCgen}
\end{equation}%
As $t\rightarrow \infty $, this asymptotically approaches a constant
\begin{equation}
\frac{d\mathcal{C}_{gen}}{dt}=\lim_{t\rightarrow \infty }\Sigma _{d-1}\sqrt{%
U(r_{min})}=\Sigma _{d-1}\sqrt{U(r_{f})},  \label{dlCgen}
\end{equation}%
where $r=r_{f}$ is the radius of the locally maximal effective potential.
Note that in the whole time evolution, we assume that the maximal extremal
hypersurface changes continuously, without any sudden jumps \cite%
{Hashimoto:2021umd}.

\begin{figure}[tbp]
\centering
\includegraphics[scale=1]{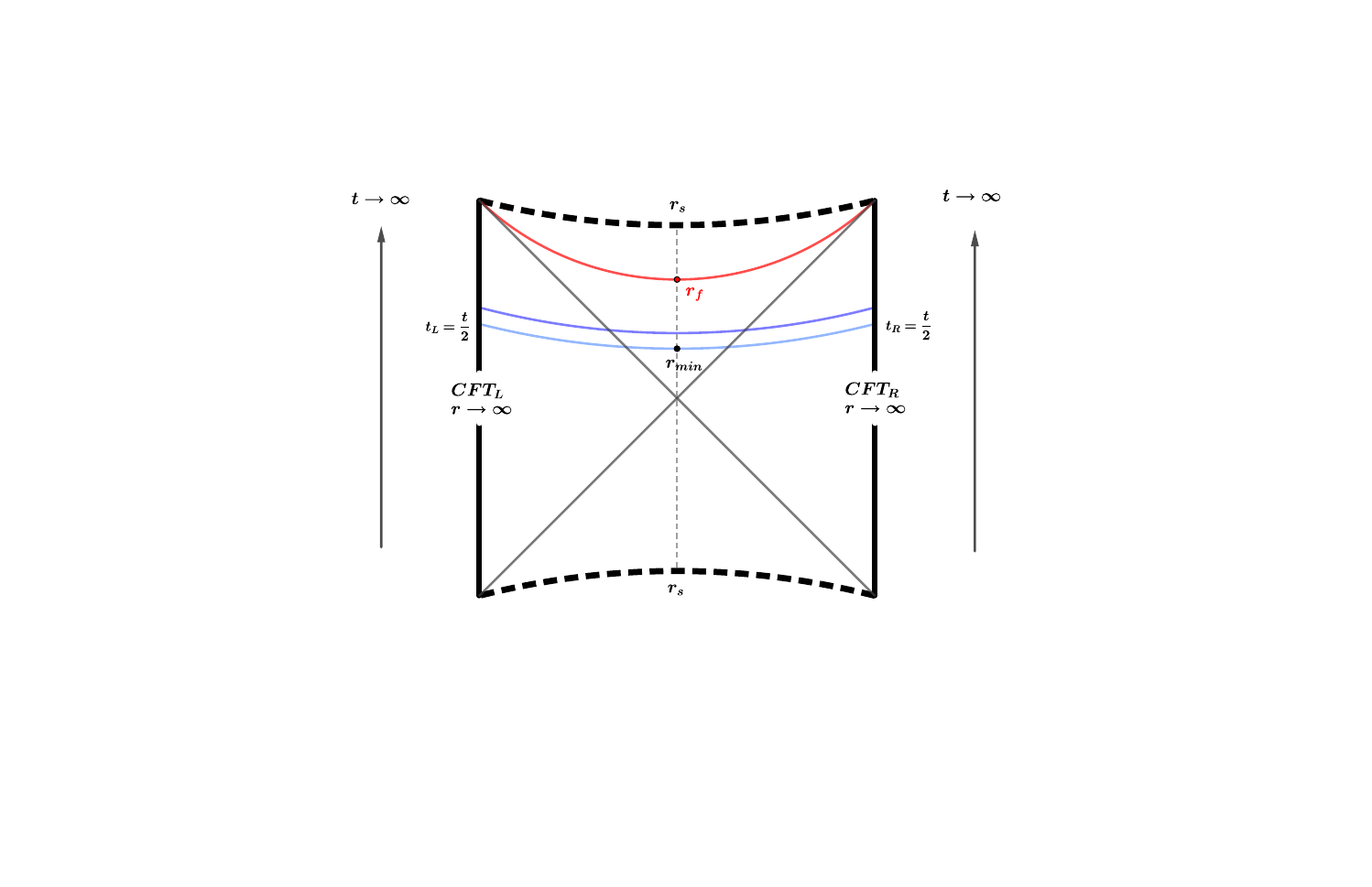}
\caption{The time evolution of the extremal hypersurface in a two-sided AdS
black hole. As time grows, the hypersurface moves up. The two nearby blue
curves correspond to the hypersurfaces anchored at two adjacent times. As
time tends to infinity, the hypersurface with minimal radius $r_{min}$
approaches to the one with $r_{f}$ (red curve), at which an effective
potential is locally maximal.}
\label{fig:PenCV}
\end{figure}

\subsubsection{Reconstruction method}

Setting $\omega =r^{d}$, we rewrite \eqref{sttgen} and \eqref{dCgen} as
\begin{equation}
t=-\frac{2}{d}\int_{\omega _{min}}^{\infty }\frac{\sqrt{U(\omega _{min})}%
\,\omega ^{\frac{1}{d}-1}}{f(\omega )\sqrt{U(\omega _{min})-U(\omega )}}%
d\omega ,  \label{stgen}
\end{equation}%
\begin{equation}
\frac{d\mathcal{C}_{gen}}{dt}=\Sigma _{d-1}\sqrt{U(\omega _{min})},
\label{dCCgen}
\end{equation}%
where%
\begin{equation}
U(\omega )=-f(\omega )\,a^{2}(\omega )\,\omega ^{2(1-\frac{1}{d})},
\label{UUU}
\end{equation}%
and
\begin{equation}
a(\omega )=1+\mathcal{K}\frac{d-2}{d}{\frac{1}{{\omega ^{\frac{4}{d}}}}}\,{%
\left\{ 2f(\omega )+\omega \,d\,\left[ \omega \,d\,f^{\prime \prime }(\omega
)+(d-3)f^{\prime }(\omega )\right] \right\} }.  \label{aw}
\end{equation}%
To proceed, we assume that $U(\omega )$ is monotonic from $\omega _{min}$ to
$\infty $. Then $U=U(\omega )$ can be inverted to a single-valued function $%
\omega =\omega (U)$. This allows us to change \eqref{stgen} as
\begin{equation}
\frac{t(U_{min})}{\sqrt{U_{min}}}=\int_{U_{min}}^{-\infty }{\frac{y(U)}{%
\sqrt{U_{min}-U}}}dU\,,  \label{stttgen}
\end{equation}%
where we have defined a function
\begin{equation}
y(U)=\frac{2}{d}\frac{\,\,a^{2}(\omega )\,\,\omega ^{1-\frac{1}{d}}}{U}\frac{%
d\omega }{dU}.  \label{y}
\end{equation}%
One can interpret \eqref{stttgen} as the Cauchy principal value to deal with
the integrand of \eqref{stgen} which blows up near the horizon $\omega _{h}$%
. Concretely, we rewrite \eqref{stttgen} as
\begin{equation}
\frac{t(U_{min})}{\sqrt{U_{min}}}=\left( \int_{U_{min}}^{\epsilon
}+\int_{\epsilon }^{-\epsilon }+\int_{-\epsilon }^{-\infty }\right) \frac{%
y(U)}{\sqrt{U_{min}-U}}dU\,,  \label{sttttgen}
\end{equation}%
where $U_{min}=U(\omega _{min})$. We assume that $\epsilon $ is small enough
that the second integral $\int_{\epsilon }^{-\epsilon }$ cancels by itself.
Then we have
\begin{equation}
\frac{t(U_{min})}{\sqrt{U_{min}}}=\left( \int_{U_{min}}^{\epsilon
}+\int_{-\epsilon }^{-\infty }\right) \frac{y(U)}{\sqrt{U_{min}-U}}dU,
\label{stttttgen}
\end{equation}%
which can be further recast as
\begin{equation}
W(U_{min})\equiv -\frac{t(U_{min})}{\sqrt{U_{min}}}-\int_{-\infty
}^{-\epsilon }\frac{y(U)}{\sqrt{U_{min}-U}}dU=\int_{\epsilon }^{U_{min}}%
\frac{y(U)}{\sqrt{U_{min}-U}}dU.  \label{sttttttgen}
\end{equation}%
Using eq. \eqref{sol}, the above integral equation can be solved
\begin{equation}
y(U)=\frac{1}{\pi }\frac{d}{dU}\int_{\epsilon }^{U}\frac{W(U_{min})}{\sqrt{%
U-U_{min}}}dU_{min}.  \label{sol st}
\end{equation}%
Combining eq. \eqref{y}, we find
\begin{equation}
\frac{2}{d}\frac{\,\,a^{2}(\omega )\,\,\omega ^{1-\frac{1}{d}}}{U}\frac{%
d\omega }{dU}=\frac{1}{\pi }\frac{d}{dU}\int_{\epsilon }^{U}\frac{W(U_{min})%
}{\sqrt{U-U_{min}}}dU_{min}.  \label{sol sst}
\end{equation}

Some remarks are in order. Firstly, our main data is the derivative of the
generalized volume. Using eq. (\ref{dCCgen}), one can calculate the function
$t(U_{min})$. In addition, we assume that the metric outside the horizon is
given. As a result, we can obtain the function $W(U_{min})$. Secondly,
consider the CV proposal where the factor $a=1$. One can see that eq. (\ref%
{sol sst}) is reduced to a first-order ordinary differential equation (ODE)
of the function $\omega (U)$. Solving the ODE and using eq. (\ref{UUU}), the
metric $f(\omega )$ can be reconstructed \cite{Hashimoto:2021umd}. Thirdly,
for a general factor, we can insert eq. (\ref{UUU}) and eq. (\ref{aw}) into
eq. (\ref{sol sst}), which is changed into a complex third-order ODE of the
metric function. Fortunately, it still can be solved by the canned
differential equation solvers such as Mathematica's NDSolve with the method
option \textquotedblleft EquationSimplification"$\rightarrow $%
\textquotedblleft Residual" \cite{wolfram}.

\subsubsection{Example}

\begin{figure}[tbp]
\centering
~~~~~~~~~~~~\includegraphics[scale=0.7]{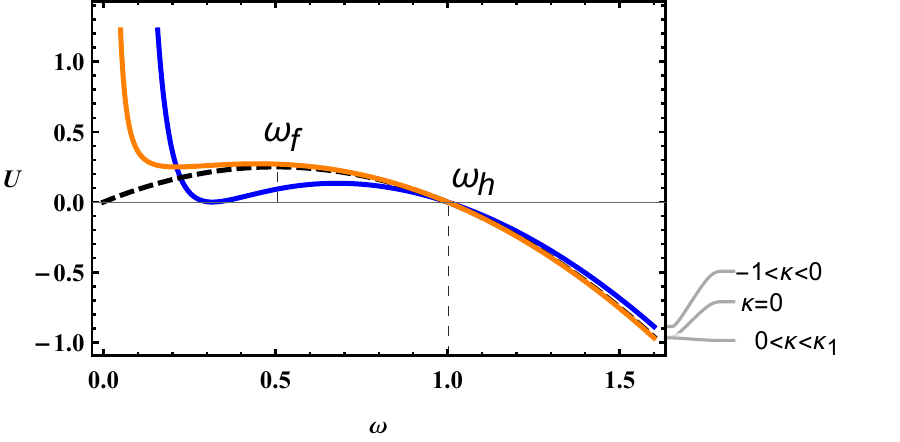}
\caption{The effective potential $U(\protect\omega )$ with $-1<\protect%
\kappa <\protect\kappa _{1}$ has the extremal point located at $\protect%
\omega _{f}$ which is within the horizon located at $\protect\omega _{h}=1$.
$U(\protect\omega )$ is monotonic from $\protect\omega _{f}$ to $\infty $.}
\label{fig:Uw}
\end{figure}

\begin{figure}[tbp]
\centering
\includegraphics[scale=0.45]{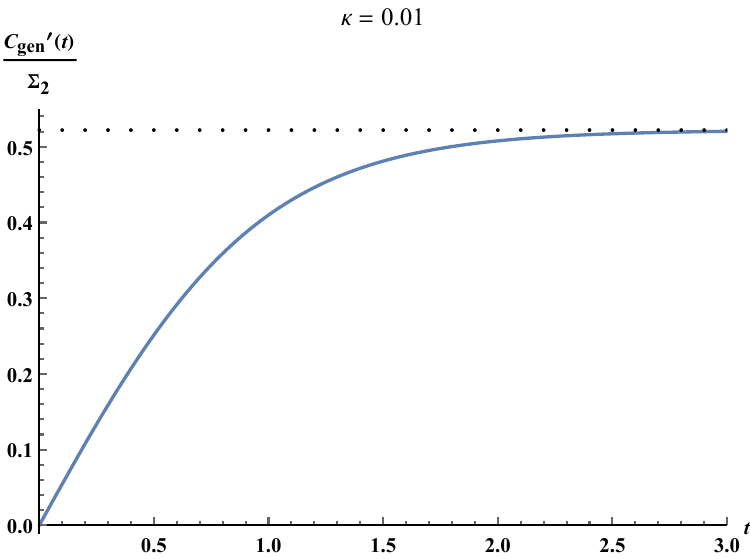} \hspace{0.2in} %
\includegraphics[scale=0.45]{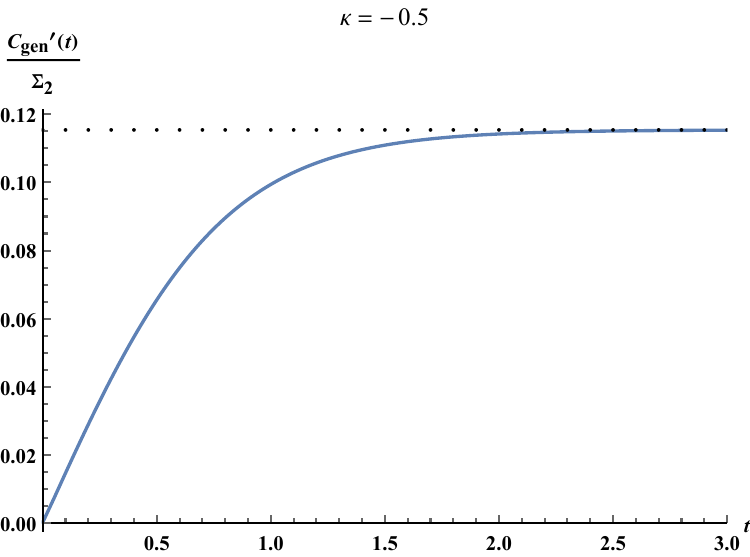}
\caption{The change rate of generalized volume (blue solid line). At the
late time, it will hit the gray dotted line, which can be described by eq.
\eqref{dlCgen}. The left and right correspond to $\protect\kappa =0.01$ and $%
\protect\kappa =-0.5$ respectively.}
\label{fig:dC}
\end{figure}

\begin{figure}[tbp]
\centering
\includegraphics[scale=0.45]{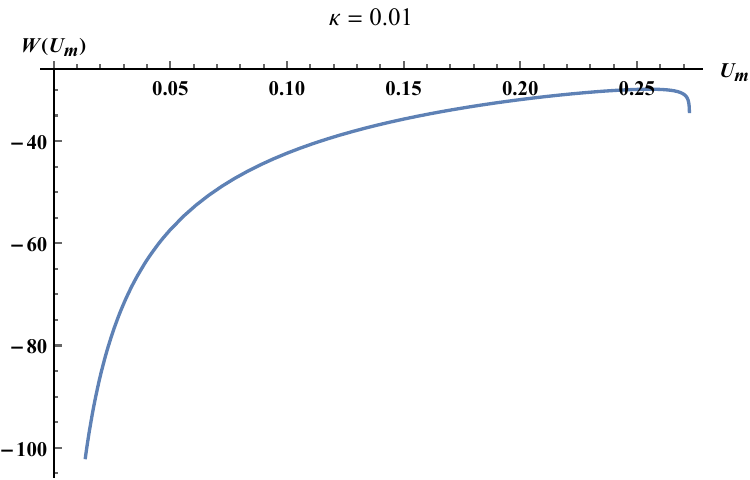} \hspace{0.2in} %
\includegraphics[scale=0.45]{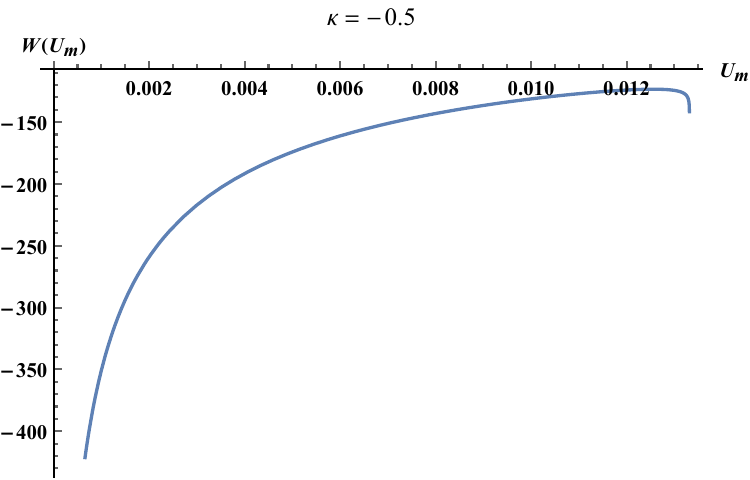}
\caption{ The function $W(U_{min})$ that is required for the reconstruction.}
\label{fig:dW}
\end{figure}

\begin{figure}[tbp]
\centering
\!\!\!\!\!\!\includegraphics[scale=0.46]{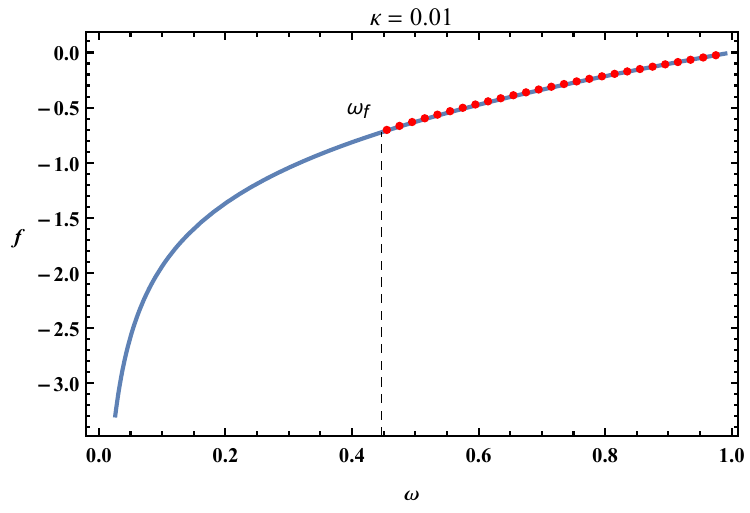} \hspace{0.3in} \!\!\!\!%
\includegraphics[scale=0.46]{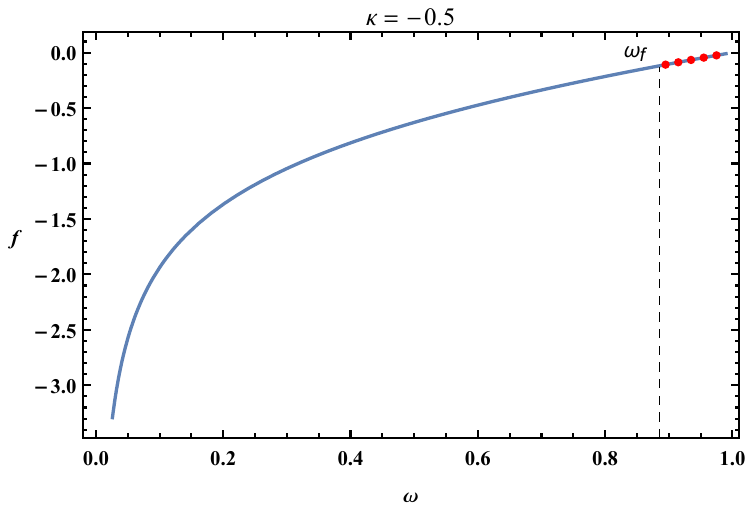}
\caption{The reconstructed metric (red dotted line) using the generalized
volume. The blue solid line is the target metric. The region deeper than $%
\protect\omega _{f}$ cannot be probed.}
\label{fig:RUw}
\end{figure}

Consider the AdS black hole with the target metric $f(\omega )=(\omega
-1)\,\omega ^{\frac{2}{d}-1}$. We read the factor (\ref{aw}) and the
potential (\ref{UUU}) as
\begin{equation}
a(\omega )=1+\kappa \left( \frac{1}{\omega }\right) ^{2},  \label{aaa}
\end{equation}%
\begin{equation}
U(\omega )=-\frac{(\omega -1)(\omega ^{2}+\kappa )^{2}}{\omega ^{3}},
\label{UUUU}
\end{equation}%
where $\kappa =d(d-1)^{2}(d-2)\mathcal{K}$. By analyzing the function (\ref%
{UUUU}), it has been pointed out that there is one local maximum inside the
horizon when $-1<\kappa <\kappa _{1}=\frac{1}{8}(47-13\sqrt{13})\approx
0.016 $ \cite{Belin:2021bga}, see figure \ref{fig:Uw}. Note that the
existence of at least one local maximum inside the horizon is the
requirement for the linear growth of the generalized volume.
In the following, we will focus on the cases with $d=3$ and $\kappa
=-0.5,0.01$. Using the target metric, we generate the data $C_{gen}^{\prime
}(t)$ and the function $W(U_{min})$ we need\footnote{%
We specify the small quantity in eq. (\ref{sttttttgen}) as $\epsilon
=10^{-9} $.}, see figures \ref{fig:dC} and \ref{fig:dW}. Solving the
third-order ODE mentioned above with the boundary condition given by the
target metric near the horizon, we reconstruct the metric inside the
horizon, see figure \ref{fig:RUw}. One can find that we build the metric
within ($\omega _{f},\omega _{h}$) where $\omega _{f}$ depends on $\kappa $.
In general, it is not possible to probe the singularity at $\omega _{s}$ if $%
\omega _{f}>\omega _{s}$.

\section{Comparison}

\label{Sec:4} Our work is closely related to refs. \cite%
{Bilson:2010ff,Hashimoto:2020mrx,Hashimoto:2021umd}, where the metric
reconstruction has been studied using three measures: EE, WL and CV. Now we
have extended their work to MI, EoP, CV2.0 and CGV. In table \ref{comp}, we
compare these seven reconstruction methods from six aspects.

1) Is the measure sensitive or insensitive to UV cutoff?

2) Is the change rate of the measure sensitive or insensitive to UV cutoff?

3) Is the measure sufficient or insufficient to reconstruct the metric?

4) Does the measure probe the exterior or interior of a black hole?

5) Does the measure depend on the spatial or temporal boundary scale?

6) Does the measure probe the spacetime in a local or non-local way?

\begin{table}[h]
\centering
\begin{tabular}{|c|c|c|c|c|c|c|}
\hline
Property & 1 & 2 & 3 & 4 & 5 & 6 \\ \hline
EE & Sensitive & Insensitive & Sufficient & Exterior & Spatial & Local \\
\hline
WL & Sensitive & Insensitive & Sufficient & Exterior & Spatial & Local \\
\hline
MI & Insensitive & Insensitive & Insufficient & Exterior & Spatial & Local
\\ \hline
EoP & Insensitive & Insensitive & Insufficient & Exterior & Spatial & Local
\\ \hline
CV & Sensitive & Insensitive & Insufficient & Interior & Temporal & Local \\
\hline
CV2.0 & Sensitive & Insensitive & Sufficient & Interior & Temporal & Local
\\ \hline
CGV & Sensitive & Insensitive & Insufficient & Interior & Temporal & Local
\\ \hline
\end{tabular}%
\caption{Comparison of reconstruction methods using seven measures across
six properties.}
\label{comp}
\end{table}

Some remarks on the comparison are in order.

i) Most of measures are sensitive to UV cutoff but their derivatives are not.

ii) For EE, WL, and CV2.0, each of them is sufficient to reconstruct the
metric, given some assumptions about spacetime symmetry and structure. For
other measures, the metric (or equivalent information) must be given for
part of the spacetime.

iii) The measures related to the EE with spatial scales probe the exterior
of black holes, while the measures related to the complexity with temporal
scales probe the interior. This reflects the complementary nature of spatial
entanglement and time-evolved complexity in encoding spacetime.

iv) All these measures probe the spacetime in a local way: reconstructing
the metric in different radial positions requires the information at
different boundary scales\footnote{%
Note that the meaning of local way defined here is slightly different from
that given in \cite{Yan:2020wcd}.}.

In order to better understand the last point, we will review briefly the
reconstruction method by the DL from shear-viscosity data \cite{Yan:2020wcd}%
. One can find that the holographic renormalization group flow of the shear
viscosity is described by the first-order ODE \cite{Yan:2020wcd}
\begin{equation}
\left( \eta -\frac{i\omega }{f}\right) ^{\prime }+\frac{i\omega }{f}\left[
z^{2}\left( \eta -\frac{i\omega }{z}\right) ^{2}-\frac{1}{z^{2}}\right] =0,
\label{eq:sec 4 sv}
\end{equation}%
where the metric ansatz \eqref{eq:sec 3 ansatz} has been used. Given the
metric $f(z)$ and the regular condition at horizon $\eta \left( \omega
,1\right) =1+i\omega $, the ODE can be solved to yield $\eta \left( \omega
,0\right) $, which is the frequency-dependent shear viscosity on the
boundary field theory. In ref. \cite{Yan:2020wcd}, a discretized
representation of the ODE \eqref{eq:sec 4 sv} is provided through a deep
neural network, where the metric is encoded as trainable weights. Given the
existence of the horizon and guided by the smoothness of spacetime, it is
shown that three typical black hole metrics can be extracted with high
accuracy from $\eta \left( \omega ,0\right) $ using DL. In particular, the
deep neural network exhibits an excellent generalization ability, indicating
that the complete metric can be well learned from the data with narrow
frequency range.

\begin{figure}[tbp]
\centering
\!\!\!\includegraphics[scale=0.7]{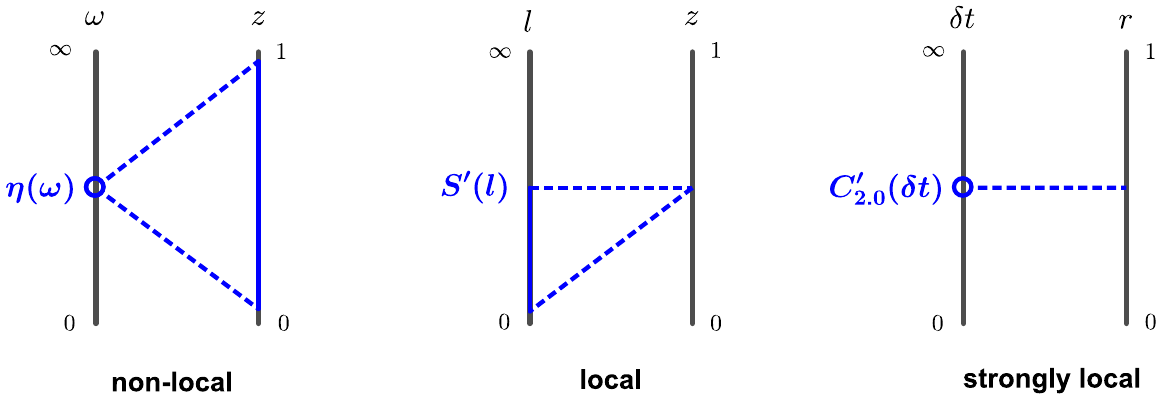}
\caption{Three typical ways to probe the spacetime that can be described as
non-local, local and strongly local, respectively. Left: The complete metric
from $z=0$ to $z=1$ can be learned from the data $\protect\eta\left(\protect%
\omega\right)$ with narrow frequency range. Middle: To build the metric $%
f(z) $ at a given $z$, the data $S^{\prime }(l)$ with a finite range $%
(0,l]$ is required. Right: The metric $f(r)$ at a given $r$ can be built
from the data $C^{\prime }_{2.0}(\protect\delta t)$ around the corresponding
$\protect\delta t$.}
\label{fig:local}
\end{figure}

\begin{figure}[tbp]
\centering
\includegraphics[scale=0.7]{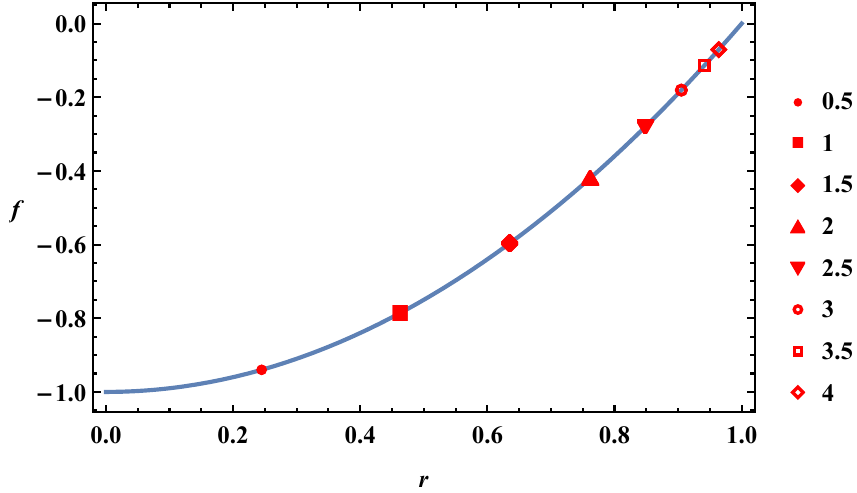}
\caption{The bulk metric reconstruction with strong locality. The red labels
denote the reconstruction results using the data $\mathcal{C}_{2.0}^{\prime
}(\protect\delta t)$ around given $\protect\delta t $. The blue solid line
is the target metric. We choose eight $\protect\delta t$ which are listed on
the right. The small ranges around them are fixed as $[\protect\delta t-0.1%
\protect\delta t,\protect\delta t+0.1\protect\delta t]$. Within each small
range, we take a uniform sample of 100 points as our data.}
\label{fig:Cf}
\end{figure}

We can classify all the reconstruction methods mentioned above into three
types, the basis of which can be seen in figure \ref{fig:local} and
explained below. The first is shear viscosity, which is characterised by the
excellent generalization ability and can be understood intuitively as
follows. Looking at eq. (\ref{eq:sec 4 sv}), one can find that in order to
determine $\eta \left( \omega ,0\right) $ at any non-zero frequency, the
complete metric from $z=0$ to $z=1$ must be known. Conversely, each $\eta
\left( \omega ,0\right) $ at a non-zero frequency contains some information
about the metric at any $z$. Therefore, if the technique of information
mining is powerful enough, it is in principle possible to reconstruct the
complete metric from a narrow frequency range, at least when the metric
function is simple enough. The second involves EE, WL, MI, EoP, CV and CGV.
Taking EE as an example, two equations (\ref{eq:sec 2 dSdl}) and (\ref%
{eq:sec 2 f(z)}) show that if we want to probe deeper into spacetime%
\footnote{%
We consider that the horizon is deeper than the boundary but shallower than
the singularity.}, we need data $S^{\prime }(l)$ at larger $l$. In addition,
to build $f(z)$ at a given $z$, the data $S^{\prime }(l)$ with a finite
range $(0,l]$ is required. The third is CV2.0. From (\ref{subsec:1 ct})-(\ref%
{subsec:1 kx4}) and the left panel of figure \ref{fig:PenCV2}, we know that
it probes the shallower spacetime as the data $\mathcal{C}_{2.0}^{\prime
}(\delta t)$ with larger $\delta t$ is input. Moreover, one can reconstruct
the metric $f(r)$ at certain $r$ \emph{if and only if} a small range of data
$\mathcal{C}_{2.0}^{\prime }(\delta t)$ around the corresponding $\delta t$
is provided, see figure \ref{fig:Cf}. We therefore argue that the
reconstruction method using CV2.0 has a certain strong locality\footnote{
Here the strong locality can be understood as a property of the mapping between the boundary data and
the bulk radial metric. In figure \ref{fig:local}, three types of mappings have been visualized. Their properties should not be confused with those of the boundary quantities themselves. For example, the complexity on the boundary is defined over the entire time slice and therefore it is highly non-local.}. Despite its
distinctiveness, CV2.0 still shares some commonality with the second type:
reconstructing metrics of different radial positions always requires the
data of different boundary scales. Note that here the measures can be
relevant to entanglement or complexity, and the boundary scale can be either
spatial or temporal.

\section{Summary}

\label{Sec:5}

Following the spirit of `It from Qubit' \cite{simons}, we investigate the
bulk reconstruction of black hole metrics using various quantum information
measures on the boundary field theories. We propose several reconstruction
methods all of which are free of UV divergence. One can see that different
mathematical ingredient appears in each method: analytic formulas for EE,
iterative data generation for MI, interpolation-generated test metrics for
EoP, simple derivatives for CV2.0, and a complex three-order ODE for CGV. We
also witness the complementarity of entanglement and complexity, and analyse
the differences and similarities among different probing ways\footnote{
Our analysis is limited to the current settings. For example, we focus on the simple strip entangling region and do not consider other shapes.}. Our results
would enrich the understanding of how field theory encodes gravity.

In the future, it would be important to reduce the assumptions of the
reconstruction methods and thus broaden their applicability. In this regard,
one can draw lessons from the bulk reconstruction that do not rely on
spacetime symmetries \cite{Bao:2019bib,Cao:2020uvb}, as well as the
illumination on entanglement shadows \cite{Hubeny1306,Bala1406,Freivogel1412}%
. Meanwhile, one can collect experimental and simulated data on quantum
information measures in strongly coupled field theories with gravity duals.
It would be intriguing to explore whether the reconstruction methods can
extract reasonable holographic spacetimes from these data\footnote{%
After this work was completed, we are aware of an exciting recent progress
\cite{Jokela2304}, where the holographic bulk metrics are reconstructed from
the derivative of EE in the lattice Yang-Mills theory.}.

\section*{Acknowledgments}

We thank Xian-Hui Ge, Yu Tian and Run-Qiu Yang for helpful discussions. This
work was supported partially by NSFC grants (No.11675097).

\bigskip



\begin{thebibliography}{999}
\bibitem{Maldacena:1997re} J.M. Maldacena, \emph{The Large N limit of
superconformal field theories and supergravity}, \href{https://www.intlpress.com/site/pub/files/_fulltext/journals/atmp/1998/0002/0002/ATMP-1998-0002-0002-a001.pdf}%
{\emph{Adv. Theor. Math. Phys.} \textbf{2} (1998) 231}[\href{https://arxiv.org/abs/hep-th/9711200}%
{\texttt{hep-th/9711200}}].


\bibitem{Gubser:1998bc} S.S.~Gubser, I.R.~Klebanov and A.M.~Polyakov, \emph{%
Gauge theory correlators from noncritical string theory}, \href{https://www.sciencedirect.com/science/article/abs/pii/S0370269398003773?via\%3Dihub}%
{\emph{Phys. Lett. B} \textbf{428} (1998) 105} [\href{https://arxiv.org/abs/hep-th/9802109}%
{\texttt{hep-th/9802109}}].


\bibitem{Witten:1998qj} E.~Witten, \emph{Anti-de Sitter space and holography}%
, \href{https://www.intlpress.com/site/pub/pages/journals/items/atmp/content/vols/0002/0002/a002/}%
{\emph{Adv. Theor. Math. Phys.} \textbf{2} (1998) 253 }[\href{https://arxiv.org/abs/hep-th/9802150}%
{\texttt{hep-th/9802150}}].


\bibitem{Susskind:1998dq} L.~Susskind and E.~Witten, \emph{The Holographic
bound in anti-de Sitter space, } (1998) [\href{https://arxiv.org/abs/hep-th/9805114}%
{\texttt{hep-th/9805114}}].


\bibitem{Liu:2020rrn} H.~Liu and J.~Sonner, \emph{Quantum many-body physics
from a gravitational lens}, \href{https://www.nature.com/articles/s42254-020-0225-1}%
{\emph{Nature Rev. Phys.} \textbf{2} (2020) 615} [\href{https://arxiv.org/abs/2004.06159}%
{\texttt{arXiv:2004.06159}}].


\bibitem{Hamilton:2006az} A.~Hamilton, D.N.~Kabat, G.~Lifschytz and
D.A.~Lowe, \emph{Holographic representation of local bulk operators, } \href{https://journals.aps.org/prd/abstract/10.1103/PhysRevD.74.066009}%
{\emph{Phys. Rev. D} \textbf{74} (2006) 066009} [\href{https://arxiv.org/abs/hep-th/0606141}%
{\texttt{hep-th/0606141}}].


\bibitem{DeJonckheere:2017qkk} T.~De Jonckheere, \emph{Modave lectures on
bulk reconstruction in AdS/CFT,} \href{https://www.researchgate.net/publication/321210795_Modave_lectures_on_bulk_reconstruction_in_AdSCFT}%
{\emph{PoS} \textbf{Modave2017} (2018) 005} [\href{https://arxiv.org/abs/1711.07787}%
{\texttt{arXiv:1711.07787}}].


\bibitem{Harlow:2018fse} D.~Harlow, \emph{TASI Lectures on the emergence of
bulk physics in AdS/CFT, }\href{https://dspace.mit.edu/handle/1721.1/121453}{%
\emph{PoS} \textbf{TASI2017} (2018) 002} [\href{https://arxiv.org/abs/1802.01040}%
{\texttt{arXiv:1802.01040}}].

\bibitem{Kajuri2003} N.~Kajuri, \emph{Lectures on bulk reconstruction, }
\href{https://doi.org/10.21468/SciPostPhysLectNotes.22}{\emph{SciPost Phys.
Lect. Notes} 22 (2021)} [\href{https://arxiv.org/abs/2003.00587}{\texttt{%
arXiv:2003.00587}}]. 


\bibitem{deHaro:2000vlm} S.~de Haro, S.N.~Solodukhin and K.~Skenderis, \emph{%
Holographic reconstruction of space-time and renormalization in the AdS /
CFT correspondence}, \href{https://link.springer.com/article/10.1007/s002200100381}%
{\emph{Commun. Math. Phys.} \textbf{217} (2001) 595} [\href{https://arxiv.org/abs/hep-th/0002230}%
{\texttt{hep-th/0002230}}].


\bibitem{Hammersley:2006cp} J.~Hammersley, \emph{Extracting the bulk metric
from boundary information in asymptotically AdS spacetimes} \href{https://iopscience.iop.org/article/10.1088/1126-6708/2006/12/047}%
{\emph{JHEP} \textbf{12} (2006) 047} [\href{https://arxiv.org/abs/hep-th/0609202}%
{\texttt{hep-th/0609202}}].


\bibitem{Hubeny:2006yu} V.E.~Hubeny, H. Liu and M.~Rangamani, \emph{%
Bulk-cone singularities \& signatures of horizon formation in AdS/CFT} \href{https://iopscience.iop.org/article/10.1088/1126-6708/2007/01/009}%
{\emph{JHEP} \textbf{01} (2007) 009} [\href{https://arxiv.org/abs/hep-th/0610041}%
{\texttt{hep-th/0610041}}].


\bibitem{Hammersley:2007ab} J.~Hammersley, \emph{Numerical metric extraction
in AdS/CFT, } \href{https://link.springer.com/article/10.1007/s10714-007-0564-6}%
{\emph{Gen. Rel. Grav.} \textbf{40} (2008) 1619} [\href{https://arxiv.org/abs/0705.0159}%
{\texttt{arXiv:0705.0159}}].


\bibitem{Bilson:2008ab} S.~Bilson, \emph{Extracting spacetimes using the
AdS/CFT conjecture, } \href{https://iopscience.iop.org/article/10.1088/1126-6708/2008/08/073}%
{\emph{JHEP} \textbf{08} (2008) 073} [\href{https://arxiv.org/abs/0807.3695}{
\texttt{arXiv:0807.3695}}].


\bibitem{Bilson:2010ff} S.~Bilson, \emph{Extracting Spacetimes using the
AdS/CFT Conjecture: Part II, } \href{https://link.springer.com/article/10.1007/JHEP02(2011)050}%
{\emph{JHEP} \textbf{02} (2011) 050} [\href{https://arxiv.org/abs/1012.1812}{
\texttt{arXiv:1012.1812}}].


\bibitem{Hubeny:2012ry} V.E.~Hubeny, \emph{Extremal surfaces as bulk probes
in AdS/CFT, } \href{https://link.springer.com/article/10.1007/JHEP07(2012)093}%
{\emph{JHEP} \textbf{07} (2012) 093} [\href{https://arxiv.org/abs/1203.1044}{%
\texttt{arXiv:1203.1044}}].


\bibitem{Balasubramanian:2013lsa} V.~Balasubramanian, B.D.~Chowdhury,
B.~Czech, J.~de Boer and M.P.~Heller, \emph{Bulk curves from boundary data
in holography, } \href{https://journals.aps.org/prd/abstract/10.1103/PhysRevD.89.086004}%
{\emph{Phys. Rev. D} \textbf{89} (2014) 086004} [\href{https://arxiv.org/abs/1310.4204}%
{\texttt{arXiv:1310.4204}}].


\bibitem{Myers:2014jia} R.C.~Myers, J.~Rao and S.~Sugishita, \emph{%
Holographic Holes in Higher Dimensions, } \href{https://link.springer.com/article/10.1007/JHEP06(2014)044}%
{\emph{JHEP} \textbf{06} (2014) 044} [\href{https://arxiv.org/abs/1403.3416}{%
\texttt{arXiv:1403.3416}}].


\bibitem{Czech:2014ppa} B.~Czech and L.~Lamprou, \emph{Holographic
definition of points and distances, } \href{https://journals.aps.org/prd/abstract/10.1103/PhysRevD.90.106005}%
{\emph{Phys. Rev. D} \textbf{90} (2014) 106005} [\href{https://arxiv.org/abs/1409.4473}%
{\texttt{arXiv:1409.4473}}].


\bibitem{Engelhardt:2016wgb} N.~Engelhardt and G.T.~Horowitz, \emph{Towards
a Reconstruction of General Bulk Metrics, } \href{https://iopscience.iop.org/article/10.1088/1361-6382/34/1/015004}%
{\emph{Class. Quant. Grav.} \textbf{34} (2017) 015004} [\href{https://arxiv.org/abs/1605.01070}%
{\texttt{arXiv:1605.01070}}].


\bibitem{Engelhardt:2016crc} N.~Engelhardt and G.T.~Horowitz, \emph{%
Recovering the spacetime metric from a holographic dual, } \href{https://www.intlpress.com/site/pub/pages/journals/items/atmp/content/vols/0021/0007/a002/}%
{\emph{Adv. Theor. Math. Phys.} \textbf{21} (2017) 1635} [\href{https://arxiv.org/abs/1612.00391}%
{\texttt{arXiv:1612.00391}}].

\bibitem{Roy:2018ehv} S.R.~Roy and D.~Sarkar, \emph{Bulk metric
reconstruction from boundary entanglement, } \href{https://journals.aps.org/prd/abstract/10.1103/PhysRevD.98.066017}%
{\emph{Phys. Rev. D} \textbf{98} (2018) 066017} [\href{https://arxiv.org/abs/1801.07280}%
{\texttt{arXiv:1801.07280}}].


\bibitem{Kabat:2018smf} D.~Kabat and G.~Lifschytz, \emph{Emergence of
spacetime from the algebra of total modular Hamiltonians, } \href{https://www.semanticscholar.org/paper/Emergence-of-spacetime-from-the-algebra-of-total-Kabat-Lifschytz/95bb021c88644d8db273cf75d8e4a4d677ac8669}%
{\emph{JHEP} \textbf{05} (2019) 017} [\href{https://arxiv.org/abs/1812.02915}%
{\texttt{arXiv:1812.02915}}].

\bibitem{Hashimoto:2020mrx} K.~Hashimoto, \emph{Building bulk from Wilson
loops, }\href{https://www.researchgate.net/publication/347642346_Building_bulk_from_Wilson_loops}%
{\emph{PTEP} \textbf{2021} (2021) 023B04} [\href{https://arxiv.org/abs/2008.10883}%
{\texttt{arXiv:2008.10883}}].

\bibitem{Caron-Huot2211} S. Caron-Huot, \emph{Holographic cameras: An eye
for the bulk, \href{https://link.springer.com/article/10.1007/JHEP03(2023)047}%
{\emph{JHEP} \textbf{03} (2023) 047}} [\href{https://arxiv.org/abs/2211.11791}%
{\texttt{arXiv:}2211.11791}].


\bibitem{Maldacena:2001kr} J.M.~Maldacena, \emph{Eternal black holes in
anti-de Sitter, }\href{https://iopscience.iop.org/article/10.1088/1126-6708/2003/04/021}%
{\emph{JHEP} \textbf{04} (2003) 021} [\href{https://arxiv.org/abs/hep-th/0106112}%
{\texttt{hep-th/0106112}}].


\bibitem{Ryu:2006bv} S.~Ryu and T.~Takayanagi, \emph{Holographic derivation
of entanglement entropy from AdS/CFT, } \href{https://journals.aps.org/prl/abstract/10.1103/PhysRevLett.96.181602}%
{\emph{Phys. Rev. Lett.} \textbf{96} (2006) 181602} [\href{https://arxiv.org/abs/hep-th/0603001}%
{\texttt{hep-th/0603001}}].


\bibitem{VanRaamsdonk:2010pw} M.~Van Raamsdonk, \emph{Building up spacetime
with quantum entanglement, }\href{https://link.springer.com/article/10.1007/s10714-010-1034-0}%
{\emph{Gen. Rel. Grav.} \textbf{42} (2010) 2323} [\href{https://arxiv.org/abs/1005.3035}%
{\texttt{arXiv:1005.3035}}].


\bibitem{Swingle:2009bg} B.~Swingle, \emph{Entanglement Renormalization and
Holography, }\href{https://journals.aps.org/prd/abstract/10.1103/PhysRevD.86.065007}%
{\emph{Phys. Rev. D} \textbf{86} (2012) 065007} [\href{https://arxiv.org/abs/0905.1317}%
{\texttt{arXiv:0905.1317}}].


\bibitem{Maldacena:2013xja} J.~Maldacena and L.~Susskind, \emph{Cool
horizons for entangled black holes, }\href{https://onlinelibrary.wiley.com/doi/10.1002/prop.201300020}%
{\emph{Fortsch. Phys.} \textbf{61} (2013) 781} [\href{https://arxiv.org/abs/1306.0533}%
{\texttt{arXiv:1306.0533}}].


\bibitem{Takayanagi T} M. Rangamani, T. Takayanagi, \emph{Aspects of
Holographic Entanglement Entropy, }\href{https://iopscience.iop.org/article/10.1088/1126-6708/2006/08/045}%
{\emph{JHEP} \textbf{08} (2018) 045} [\href{https://arxiv.org/abs/hep-th/0605073}%
{\texttt{hep-th/0605073}}].


\bibitem{Susskind:2014moa} L.~Susskind, \emph{Entanglement is not enough, }
\href{https://www.semanticscholar.org/paper/Entanglement-is-not-enough-Susskind/e8f841ff250801231bb0f423e062ec189c8492b7}%
{\emph{Fortsch. Phys.} \textbf{64} (2016) 49} [\href{https://arxiv.org/abs/1411.0690}%
{\texttt{arXiv:1411.0690}}].


\bibitem{Susskind:2014rva} L.~Susskind, \emph{Computational Complexity and
Black Hole Horizons, }\href{https://www.semanticscholar.org/paper/Computational-complexity-and-black-hole-horizons-Susskind/3d8d8010dc5c59f6321af3b81edc999130abdfa7}%
{\emph{Fortsch. Phys.} \textbf{64} (2016) 24} [\href{https://arxiv.org/abs/1403.5695}%
{\texttt{arXiv:1403.5695}}].


\bibitem{Hashimoto:2021umd} K.~Hashimoto and R.~Watanabe, \emph{Bulk
reconstruction of metrics inside black holes by complexity, } \href{https://www.researchgate.net/publication/354840827_Bulk_reconstruction_of_metrics_inside_black_holes_by_complexity}%
{\emph{JHEP} \textbf{09} (2021) 165} [\href{https://arxiv.org/abs/2103.13186}%
{\texttt{arXiv:2103.13186}}].


\bibitem{Skenderis0209} K. Skenderis, \emph{Lecture notes on holographic
renormalization,} \href{https://iopscience.iop.org/article/10.1088/0264-9381/19/22/306}%
{\emph{Class. Quantum. Gravity} \textbf{19} (2002) 5849} [\href{https://arxiv.org/abs/hep-th/0209067}%
{\texttt{hep-th/0209067}}].


\bibitem{Papadimitriou2016} I. Papadimitriou, \emph{Lectures on Holographic
Renormalization,} \emph{Springer Proceedings in Physics Vol. 176:
Theoretical Frontiers in Black Holes and Cosmology}, Springer Press (2016).


\bibitem{Jokela:2020auu} N.~Jokela and A.~P\"{o}nni, \emph{Towards precision
holography, }\href{https://journals.aps.org/prd/abstract/10.1103/PhysRevD.103.026010}%
{\emph{Phys. Rev. D} \textbf{103} (2021) 026010} [\href{https://arxiv.org/abs/2007.00010}%
{\texttt{arXiv:2007.00010}}].



\bibitem{Hashimoto:2018ftp} K.~Hashimoto, S.~Sugishita, A.~Tanaka and
A.~Tomiya, \emph{Deep learning and the AdS/CFT correspondence, } \href{https://journals.aps.org/prd/abstract/10.1103/PhysRevD.98.046019}%
{\emph{Phys. Rev. D} \textbf{98} (2018) 046019} [\href{https://arxiv.org/abs/1802.08313}%
{\texttt{arXiv:1802.08313}}].


\bibitem{Hashimoto:2018bnb} K.~Hashimoto, S.~Sugishita, A.~Tanaka and
A.~Tomiya, \emph{Deep Learning and Holographic QCD, } \href{https://journals.aps.org/prd/abstract/10.1103/PhysRevD.98.106014}%
{\emph{Phys. Rev. D} \textbf{98} (2018) 106014} [\href{https://arxiv.org/abs/1809.10536}%
{\texttt{arXiv:1809.10536}}].


\bibitem{Hashimoto:2019bih} K.~Hashimoto, \emph{AdS/CFT correspondence as a
deep Boltzmann machine, }\href{https://journals.aps.org/prd/abstract/10.1103/PhysRevD.99.106017}%
{\emph{Phys. Rev. D} \textbf{99} (2019) 106017} [\href{https://arxiv.org/abs/1903.04951}%
{\texttt{arXiv:1903.04951}}].


\bibitem{Tan:2019czc} J.~Tan and C.B.~Chen, \emph{Deep learning the
holographic black hole with charge, } \href{https://www.worldscientific.com/doi/abs/10.1142/S0218271819501530}%
{\emph{Int. J. Mod. Phys. D} \textbf{28} (2019) 1950153} [\href{https://arxiv.org/abs/1908.01470}%
{\texttt{arXiv:1908.01470}}].


\bibitem{Akutagawa:2020yeo} T.~Akutagawa, K.~Hashimoto and T.~Sumimoto,
\emph{Deep Learning and AdS/QCD, } \href{https://journals.aps.org/prd/abstract/10.1103/PhysRevD.102.026020}%
{\emph{Phys. Rev. D} \textbf{102} (2020) 026020} [\href{https://arxiv.org/abs/2005.02636}%
{\texttt{arXiv:2005.02636}}].


\bibitem{Yan:2020wcd} Y.K.~Yan, S.F.~Wu, X.H.~Ge and Y.~Tian, \emph{Deep
learning black hole metrics from shear viscosity, } \href{https://journals.aps.org/prd/abstract/10.1103/PhysRevD.102.101902}%
{\emph{Phys. Rev. D} \textbf{102} (2020) 101902(R)} [\href{https://arxiv.org/abs/2004.12112}%
{\texttt{arXiv:2004.12112}}].


\bibitem{Hashimoto:2020jug} K.~Hashimoto, H.Y.~Hu and Y.Z.~You, \emph{Neural
ordinary differential equation and holographic quantum chromodynamics, }
\href{https://iopscience.iop.org/article/10.1088/2632-2153/abe527}{\emph{\
Mach. Learn. Sci. Tech.} \textbf{2} (2021) 035011} [\href{https://arxiv.org/abs/2006.00712}%
{\texttt{arXiv:2006.00712}}].


\bibitem{Hashimoto:2021ihd} K.~Hashimoto, K.~Ohashi and T.~Sumimoto, \emph{%
Deriving the dilaton potential in improved holographic QCD from the meson
spectrum, } \href{https://journals.aps.org/prd/abstract/10.1103/PhysRevD.105.106008}%
{\emph{Phys. Rev. D} \textbf{105} (2022) 106008} [\href{https://arxiv.org/abs/2108.08091}%
{\texttt{arXiv:2108.08091}}].


\bibitem{Katsube:2022ofz} R.~Katsube, W.H.~Tam, M.~Hotta and Y.~Nambu, \emph{%
Deep learning metric detectors in general relativity, } \href{https://journals.aps.org/prd/abstract/10.1103/PhysRevD.106.044051}%
{\emph{Phys. Rev. D} \textbf{106} (2022) 044051} [\href{https://arxiv.org/abs/2206.03006}%
{\texttt{arXiv:2206.03006}}].


\bibitem{Hashimoto:2022eij} K.~Hashimoto, K.~Ohashi and T.~Sumimoto,\emph{\
Deriving dilaton potential in improved holographic QCD from chiral
condensate, } \href{https://academic.oup.com/ptep/article/2023/3/033B01/7043484?login=false}%
{\emph{PTEP} \textbf{2023} (2023) 033B01} [\href{https://arxiv.org/abs/2209.04638}%
{\texttt{arXiv:2209.04638}}].


\bibitem{Li:2022zjc} K.~Li, Y.~Ling, P.~Liu and M.H.~Wu, \emph{Learning the
black hole metric from holographic conductivity, } \href{https://journals.aps.org/prd/abstract/10.1103/PhysRevD.107.066021}%
{\emph{Phys. Rev. D} \textbf{107} (2023) 066021} [\href{https://arxiv.org/abs/2209.05203}%
{\texttt{arXiv:2209.05203}}].


\bibitem{You:2017guh} Y.Z.~You, Z.~Yang and X.L.~Qi, \emph{Machine Learning
Spatial Geometry from Entanglement Features, } \href{https://journals.aps.org/prb/abstract/10.1103/PhysRevB.97.045153#:~:text=Machine\%20learning\%20is\%20a\%20fast\%20developing\%20area\%20that,study\%20of\%20the\%20holography\%20duality\%20in\%20quantum\%20gravity.}%
{\emph{Phys. Rev. B} \textbf{97} (2018) 045153} [\href{https://arxiv.org/abs/1709.01223}%
{\texttt{arXiv:1709.01223}}].


\bibitem{Hu:2019nea} H.Y.~Hu, S.H.~Li, L.~Wang and Y.Z.~You, \emph{Machine
Learning Holographic Mapping by Neural Network Renormalization Group, }
\href{https://journals.aps.org/prresearch/abstract/10.1103/PhysRevResearch.2.023369}%
{\emph{Phys. Rev. Res.} \textbf{2} (2020) 023369} [\href{https://arxiv.org/abs/1903.00804}%
{\texttt{arXiv:1903.00804}}].


\bibitem{Han:2019wue} X.~Han and S.A.~Hartnoll, \emph{Deep Quantum Geometry
of Matrices, }\href{https://journals.aps.org/prx/abstract/10.1103/PhysRevX.10.011069}%
{\emph{Phys. Rev. X} \textbf{10} (2020) 011069} [\href{https://arxiv.org/abs/1906.08781}%
{\texttt{arXiv:1906.08781}}].


\bibitem{Lam:2021ugb} J.~Lam and Y.Z.~You, \emph{Machine learning
statistical gravity from multi-region entanglement entropy, } \href{https://escholarship.org/uc/item/8h15k5mr}%
{\emph{Phys. Rev. Res.} \textbf{3} (2021) 043199} [\href{https://arxiv.org/abs/2110.01115}%
{\texttt{arXiv:2110.01115}}].


\bibitem{Wolf:2007tdq} M.M.~Wolf, F.~Verstraete, M.B.~Hastings and
J.I.~Cirac, \emph{Area Laws in Quantum Systems: Mutual Information and
Correlations, } \href{https://journals.aps.org/prl/abstract/10.1103/PhysRevLett.100.070502}%
{\emph{Phys. Rev. Lett.} \textbf{100} (2008) 070502} [\href{https://arxiv.org/abs/0704.3906}%
{\texttt{arXiv:0704.3906}}].


\bibitem{Terhal2002} B.M. Terhal, M. Horodecki, D.W. Leung and D.P.
DiVincenzo, \emph{The entanglement of purification, } \href{https://aip.scitation.org/doi/10.1063/1.1498001}%
{\emph{J. Math. Phys.} \textbf{43} (2002) 4286} [\href{https://arxiv.org/abs/quant-ph/0202044}%
{\texttt{quant-ph/0202044}}].


\bibitem{Couch:2016exn} J.~Couch, W.~Fischler and P.H.~Nguyen, \emph{Noether
charge, black hole volume, and complexity, } \href{https://link.springer.com/article/10.1007/JHEP03(2017)119}%
{\emph{JHEP} \textbf{03} (2017) 119} [\href{https://arxiv.org/abs/1610.02038}%
{\texttt{arXiv:1610.02038}}].


\bibitem{Belin:2021bga} A.~Belin, R.C.~Myers, S.M.~Ruan, G.S\'arosi and
A.J.~Speranza, \emph{Does complexity equal anything?, } \href{https://journals.aps.org/prl/abstract/10.1103/PhysRevLett.128.081602}%
{\emph{Phys. Rev. Lett.} \textbf{128} (2022) 081602} [\href{https://arxiv.org/abs/2111.02429}%
{\texttt{arXiv:2111.02429}}].


\bibitem{Poly} A.D.~Polyanin, Handbook of Integral Equations, CRC Press.
(1998)


\bibitem{Fischler:2012uv} W.~Fischler, A.~Kundu and S.~Kundu, \emph{%
Holographic mutual information at finite temperature, } \href{https://journals.aps.org/prd/abstract/10.1103/PhysRevD.87.126012}%
{\emph{Phys. Rev. D} \textbf{87} (2013) 126012} [\href{https://arxiv.org/abs/1212.4764}%
{\texttt{arXiv:1212.4764}}].


\bibitem{Swingle:2010jz} B.~Swingle, \emph{Mutual information and the
structure of entanglement in quantum field theory, }\href{https://pirsa.org/10100087}%
{\emph{October 2010}} [\href{https://arxiv.org/abs/1010.4038}{\texttt{\
arXiv:1010.4038}}]. 


\bibitem{Espindola:2018ozt} R.~Esp\'\i{}ndola, A.~Guijosa and J.F.~Pedraza,
\emph{Entanglement wedge reconstruction and entanglement of purification, }
\href{https://link.springer.com/article/10.1140/epjc/s10052-018-6140-2}{
\emph{Eur. Phys. J. C} \textbf{78} (2018) 646} [\href{https://arxiv.org/abs/1804.05855}%
{\texttt{arXiv:1804.05855}}].


\bibitem{Bagchi2015} S. Bagchi and A.K. Pati, \emph{Monogamy, polygamy, and
other properties of entanglement of purification, } \href{https://journals.aps.org/pra/abstract/10.1103/PhysRevA.91.042323}%
{\emph{Phys. Rev. A} \textbf{91} (2015) 042323} [\href{https://arxiv.org/abs/1502.01272}%
{\texttt{arXiv:1502.01272}}].


\bibitem{Caputa1812} P. Caputa, M. Miyaji, T. Takayanagi, K. Umemoto, \emph{%
Holographic entanglement of purification from conformal field theories, } \href{
https://doi.org/10.1103/PhysRevLett.122.111601}%
{\emph{	Phys. Rev. Lett.} \textbf{122} (2019) 111601} [\href{https://arxiv.org/abs/1812.05268}%
{\texttt{arXiv:1812.05268}}].



\bibitem{Takayanagi:2017knl} T.~Takayanagi and K.~Umemoto, \emph{%
Entanglement of purification through holographic duality, } \href{
https://doi.org/10.1063/1.1498001}%
{\emph{Nature Phys.} \textbf{14} (2018) 573} [\href{https://arxiv.org/abs/1708.09393}%
{\texttt{arXiv:1708.09393}}].


\bibitem{Tamaoka:2018ned} K.~Tamaoka, \emph{Entanglement Wedge Cross Section
from the Dual Density Matrix, } \href{https://journals.aps.org/prl/abstract/10.1103/PhysRevLett.122.141601}%
{\emph{Phys. Rev. Lett.} \textbf{122} (2019) 141601} [\href{https://arxiv.org/abs/1809.09109}%
{\texttt{arXiv:1809.09109}}].


\bibitem{Yang1810} R.Q. Yang, C.Y. Zhang and W.M. Li, \emph{Holographic
entanglement of purification for thermofield double states and thermal
quench, } \href{https://link.springer.com/article/10.1007/JHEP01\%282019\%29114}%
{\emph{JHEP} \textbf{01} (2019) 114} [\href{https://arxiv.org/abs/1810.00420}%
{\texttt{arXiv:1810.00420}}].

\bibitem{Watrous0804} J. Watrous, \emph{Quantum Computational Complexity, }[
\href{https://arxiv.org/abs/1810.00420}{\texttt{arXiv:0804.3401}}].


\bibitem{Brown:2015bva} A.R.~Brown, D.A.~Roberts, L.~Susskind, B.~Swingle
and Y.~Zhao, \emph{Holographic Complexity Equals Bulk Action?, } \href{https://journals.aps.org/prl/abstract/10.1103/PhysRevLett.116.191301}%
{\emph{Phys. Rev. Lett.} \textbf{116} (2016) 191301} [\href{https://arxiv.org/abs/1509.07876}%
{\texttt{arXiv:1509.07876}}].


\bibitem{Carmi:2016wjl} D.~Carmi, R.C.~Myers and P.~Rath, \emph{Comments on
Holographic Complexity, }\href{https://link.springer.com/article/10.1007/JHEP03(2017)118}%
{\emph{JHEP} \textbf{03} (2017) 118} [\href{https://arxiv.org/abs/1612.00433}%
{\texttt{arXiv:1612.00433}}].


\bibitem{Carmi:2017jqz} D.~Carmi, S.~Chapman, H.~Marrochio, R.C.~Myers and
S.~Sugishita, \emph{On the Time Dependence of Holographic Complexity, }
\href{https://link.springer.com/article/10.1007/JHEP11\%282017\%29188}{\emph{%
\ \ JHEP} \textbf{11} (2017) 188} [\href{https://arxiv.org/abs/1709.10184}{
\texttt{arXiv:1709.10184}}].


\bibitem{An:2018dbz} Y.S.~An, R.G.~Cai and Y.~Peng, \emph{Time Dependence of
Holographic Complexity in Gauss-Bonnet Gravity, } \href{https://journals.aps.org/prd/abstract/10.1103/PhysRevD.98.106013}%
{\emph{Phys. Rev. D} \textbf{98} (2018) 106013} [\href{https://arxiv.org/abs/1805.07775}%
{\texttt{arXiv:1805.07775}}].


\bibitem{wolfram} Wolfram Language and System Documentation Center. \href{https://reference.wolfram.com/language/tutorial/NDSolveDAE.html}%
{\emph{Numerical Solution of Differential-Algebraic Equations.}}


\bibitem{Omidi:2022whq} F.~Omidi, \emph{Generalized Volume-Complexity For
Two-Sided Hyperscaling Violating Black Branes, } \href{https://link.springer.com/article/10.1007/JHEP01(2023)105}%
{\emph{JHEP} \textbf{01} (2023) 105} [\href{https://arxiv.org/abs/2207.05287}%
{\texttt{arXiv:2207.05287}}].


\bibitem{Brown:2015lvg} A.R.~Brown, D.A.~Roberts, L.~Susskind, B.~Swingle
and Y.~Zhao, \emph{Complexity, action, and black holes, } \href{https://journals.aps.org/prd/abstract/10.1103/PhysRevD.93.086006}%
{\emph{Phys. Rev. D} \textbf{93} (2016) 086006} [\href{https://arxiv.org/abs/1512.04993}%
{\texttt{arXiv:1512.04993}}].


\bibitem{Burda:2018rpb} P.~Burda, R.~Gregory and A.~Jain, \emph{Holographic
reconstruction of bubble spacetimes, } \href{https://journals.aps.org/prd/abstract/10.1103/PhysRevD.99.026003}%
{\emph{Phys. Rev. D} \textbf{99} (2019) 026003} [\href{https://arxiv.org/abs/1804.05202}%
{\texttt{arXiv:1804.05202}}].


\bibitem{Hernandez-Cuenca:2020ppu} S.~Hern\'{a}ndez-Cuenca and
G.T.~Horowitz, \emph{Bulk reconstruction of metrics with a compact space
asymptotically, } \href{https://link.springer.com/article/10.1007/JHEP08(2020)108}%
{\emph{JHEP} \textbf{08} (2020) 108} [\href{https://arxiv.org/abs/2003.08409}%
{\texttt{arXiv:2003.08409}}].

\bibitem{simons} Simons foundation: \href{https://www.simonsfoundation.org/mathematics-physical-sciences/it-from-qubit/}%
{\emph{It from Qubit: Simons collaboration on quantum Fields, gravity and
information.}}

\bibitem{Bao:2019bib} N.~Bao, C.~Cao, S.~Fischetti and C.~Keeler, \emph{%
Towards Bulk Metric Reconstruction from Extremal Area Variations, } \href{https://iopscience.iop.org/article/10.1088/1361-6382/ab377f/pdf}%
{\emph{Class. Quant. Grav.} \textbf{36} (2019) 185002} [\href{https://arxiv.org/abs/1904.04834}%
{\texttt{arXiv:1904.04834}}].


\bibitem{Cao:2020uvb} C.~Cao, X.L.~Qi, B.~Swingle and E.~Tang, \emph{%
Building Bulk Geometry from the Tensor Radon Transform, } \href{https://link.springer.com/article/10.1007/JHEP12(2020)033}%
{\emph{JHEP} \textbf{12} (2020) 033} [\href{https://arxiv.org/abs/2007.00004}%
{\texttt{arXiv:2007.00004}}].


\bibitem{Hubeny1306} V.E. Hubeny, H. Maxfield, M. Rangamani and E. Tonni,
\emph{Holographic entanglement plateaux}, \href{https://link.springer.com/article/10.1007/JHEP08(2013)092}%
{\emph{JHEP} \textbf{08} (2013) 092} [\href{https://arxiv.org/abs/1306.4004}{%
\texttt{arXiv:1306.4004}}].

\bibitem{Bala1406} V. Balasubramanian, B.D. Chowdhury, B. Czech and J. de
Boer, \emph{Entwinement and the emergence of spacetime}, \href{https://link.springer.com/article/10.1007/JHEP01(2015)048}%
{\emph{JHEP} \textbf{01} (2015) 048} [\href{https://arxiv.org/abs/1406.5859}{%
\texttt{arXiv:1406.5859}}].


\bibitem{Freivogel1412} B. Freivogel, R. Jefferson, L. Kabir, B. Mosk and
I.-S. Yang, \emph{Casting shadows on holographic reconstruction}, \href{https://journals.aps.org/prd/abstract/10.1103/PhysRevD.91.086013}%
{\emph{Phys. Rev. D} \textbf{91} (2015) 086013} [\href{https://arxiv.org/abs/1412.5175}%
{\texttt{arXiv:1412.5175}}].

\bibitem{Jokela2304} N. Jokela, A. P\"{o}nni, T. Rindlisbacher, K.
Rummukainen and A. Salami, \emph{Disentangling the gravity dual of
Yang-Mills theory, }[\href{https://arxiv.org/abs/2304.08949}{\texttt{arXiv:}%
2304.08949}].
\end{thebibliography}
\end{document}